\documentclass[twocolumn,showpacs,preprintnumbers,amsmath,amssymb,prb,aps]{revtex4}

\usepackage{epsfig} %Include figure files
\usepackage{graphicx}% Include figure files
\usepackage{dcolumn}% Align table columns on decimal point
\usepackage{bm}% bold math
\usepackage{epstopdf}
\usepackage{amsmath}
\usepackage{multirow}

\begin{document}

\title{ Tuning the structural, electronic and magneto-transport properties of spin-orbit Mott insulator Sr$_{2}$IrO$_{4}$ }
\author{Priyamedha Sharma$^1$}
\author{Saurabh Singh$^2$$^,$$^3$}
\author{Kentaro Kuga$^2$$^,$$^3$}
\author{Tsunehiro Takeuchi$^2$$^,$$^3$}
\author{R. Bindu$^1$}
\altaffiliation{Corresponding author: bindu@iitmandi.ac.in}
\affiliation{$^1$School of Basic Sciences, Indian Institute of Technology Mandi, Kamand, Himachal Pradesh- 175005, India\linebreak
$^2$Toyota Technological Institute, Nagoya, Aichi 468-8511, Japan\linebreak
$^3$Japan Science and Technology Agency, Kawaguchi, Saitama 332-0012, Japan}

\date{\today}
\begin{abstract}
In this work, we investigate the tunability of the structural, electronic and magneto transport properties of polycrystalline Sr$_{2}$IrO$_{4}$. Both the compounds exhibit transition from paramagnetic to canted antiferromagnetic (AFM) structure $\sim$240 K. At low temperatures, the extent of the bifurcation of the magnetisation curves during the field cooled and zero field cooled cycles establishes that the magnetic anisotropy in the \emph{as prepared} sample is more as compared to the \emph{vacuum annealed} one. Based on the behaviours of the structural parameters and the magnetic studies, our results show that the canted AFM structure is stabilized in a larger temperature range in the case of the \emph{annealed} sample ($\sim$165K) as compared to the \emph{as prepared} one ($\sim$115K). At low temperatures, for both the samples, a phase, of glassy nature competes with the canted AFM phase. The temperature extent to which both these phases coexist for the \emph{as prepared} and the \emph{annealed one} is $\sim$115K and $\sim$70K, respectively. We also observe the importance of hybridization of the Ir 5\emph{d} and O 2\emph{p} states in the canted AFM phase. Our results show a strong link between the extent of magnetic anisotropy and the difference in the in-plane Ir-O-Ir bond angles between these two samples.The transport studies reveal that for both the samples, in the high temperature region of study, the conduction mechanism is governed by Arrhenius model. In the intermediate temperature range, variable range hopping (VRH) and Arrhenius models govern the transport in the \emph{as prepared} and the \emph{annealed ones}, respectively. At low temperatures, the conduction mechanism occurs through Efros–Shklovskii-VRH and VRH mechanisms for the \emph{as prepared} and the \emph{annealed} samples, respectively.The magneto resistance measurements indicate higher negative magneto resistance in the $\emph{annealed}$ at all temperatures which can be explained by larger reduction in  spin dependent scattering and a higher sensitivity of Ir-O-Ir in-plane bond angle to the applied magnetic field. The field dependence of magneto resistance at 10K suggests a co-existing glassy magnetic phase along with canted antiferromagnetic structure in the $\emph{as prepared}$ sample and a suppression of this glassy magnetic phase in the $\emph{annealed}$ sample. The combined analysis of all the results highlight the role of disorder for the magnetic and transport properties of this compound.

\end{abstract}


% x-ray diffraction, phase separation, magnetic oxides, strongly correlated electron systems

\maketitle

\section{Introduction}

Iridates have attracted significant amount of research due to its unconventional properties like stabilization of different topological phases of matter, superconductivity, spin liquid behaviours, kitaev modes etc.\cite{witczak2014,rau2016,cao2018} These properties emerge due to strong interplay between the spin-orbit coupling (SOC), on-site Coulombic interaction, Hund’s coupling and crystal field interactions. In the case of Ruddlesden popper series Sr$_{n+1}$B$_{n}$O$_{3n+1}$, for n=1 that represent the dimensionality, as one varies the transition metal ion, B from having electronic character 3$\emph{d}$ to 5$\emph{d}$, unique properties emerge due to strong interplay of various energy scales as mentioned above. The distinguishing factors are observation of charge gap for Sr$_{2}$IrO$_{4}$ despite the extended nature of the 5$\emph{d}$ states in comparison to the 3$\emph{d}$ states\cite{kim2008}; observation of high magnetic transition temperatures but with low moment values; the canting of the magnetic moment follows the rotation of the IrO$_{6}$ octahedra; weak coupling between the transport and magnetism; saturating resistivity at low temperatures etc.\cite{ye2013,cao2018}

Interest in Sr$_{2}$IrO$_{4}$ compound were driven initially due to its structural analogue with high Tc cuprate based superconductor from the structural, magnetic and electronic point of view.\cite{huang1994,cava1994,cao1998} Many theoretical studies predicted a possibility of superconductivity upon electron or hole doping in Sr$_{2}$IrO$_{4}$.\cite{wang2011,yang2014,gao2015} In contradiction to these results, the transport and magnetic measurements on the doped bulk samples did not show superconductivity. In this compound, electron doping can be achieved by substituting La in the place of Sr ion or by introducing oxygen vacancies.\cite{chen2015,de2015,pincini2017,gretarsson2016} The hole doping is realized by substituting Ru/Rh at Ir site or K at Sr site.\cite{klein2008,calder2015,lee2012,yuan2015} It has been observed that even a small amount of chemical doping led to the evolution of spin-orbit Mott insulating state into an exotic metallic state with the increase in the doping level. On the other hand, oxygen deficient single crystals produced by annealing in vacuum exhibited a metal to insulator transition at low temperatures.\cite{korneta2010} The nature of the metallic state induced by these electron or hole doping is a subject of interest in recent years due to their novel properties such as robust AFM metallic state,\cite{wang2018} phase separation\cite{chen2015,calder2015} etc,.

The ground states of the 5$\emph{d}$ systems are governed by the hopping of the electrons through direct d-d overlap and/or through the intervening oxygen ions. Such overlap is strongly dependent on the local structural parameters. Hence, strain and/or disorder play critical role in deciding the physical properties. One of the novel properties observed in Sr$_{2}$IrO$_{4}$ is its unusual link between its exotic properties and the structural aspects through the spin orbit coupling (SOC). At room temperature (RT) this compound stabilizes in tetragonal structure with $\emph{I41/acd}$ spacegroup.\cite{crawford1994,huang1994} The characteristics of this space group is that there occurs antiphase rotation of the corner linked IrO$_{6}$ octahedra along the layering axis and there by leading to the doubling of the c-parameters in comparison to the high temperature high symmetric structural phase (T$>$ 1000 K) with space group $\emph{I4/mmm}$.\cite{ranjbar2015} This leads to uniaxial negative thermal expansion (along the c-axis) and hence an increase in the unit cell volume. The octahedral rotation breaks the inversion center on oxygen atom connecting the two Ir atoms which leads to the appearance of Dzyaloshinskii-moriya (DM) interaction\cite{dzyaloshinsky1958} along with the super exchange interaction between the J$_{eff}$=1/2 pseudospins.\cite{jackeli2009,katukuri2014} The DM interaction induces a canting of antiferromagnetically coupled  J$_{eff}$=1/2 spins leading to the appearance of a weak ferromagnetism below $\sim$240K. The J$_{eff}$=1/2 moments track the in-plane octahedral rotation through a strong magnetoelastic coupling.\cite{ye2013} It is important to note that the structural transition from $\emph{I4/mmm}$ to $\emph{I41/acd}$ space group is of second order kind where a variation in the transition temperature is expected to occur at local scale due to disorder.\cite{rocha2020} It has also been realized that positive thermal expansion occurs in the $\emph{I4/mmm}$ phase. Such thermal tendencies due to local disorder can give rise to unconventional behaviours.

In single crystal samples of Sr$_{2}$IrO$_{4}$, the reported magnetization curves along a and c axes, exhibit distinct features at around $\sim$100K and $\sim$25K, in addition to the weak ferromagnetic transition at $\sim$240K. These features are attributed to a spin reorientation transition leading to a decrease of magnetic anisotropy along a and c axes.\cite{ge2011} Also, temperature dependent dielectric constant measurements\cite{chikara2009} shows a peak at $\sim$100K with a strong frequency dependence, indicating a frustrating magnetic order at low temperatures. There are also reports of co-existence of cluster glass phase present along with canted AFM phase at low temperatures in polycrystalline samples.\cite{manna2014} All these results indicate a complex magnetic structure at low temperatures and it fairly deviates from the canted AFM structure with moments confined to the \emph{ab}-plane.

There are contradictory reports in literature with regard to conduction mechanism in single crystal as well as polycrystalline samples. Three  temperature regions with different thermal activation behaviour is observed in single crystal specimens.\cite{ge2011} In the case of polycrystalline samples, Bhatti \emph{et al}\cite{bhatti2014} have reported 2-dimensional (2D) variable range hopping(VRH) in three different temperature regions with different localization lengths. But Kini \emph{et al}\cite{kini2006} have observed resistivity proportional to ln(T) in the paramagnetic phase and Arrhenius behaviour in the intermediate temperature range, followed by a 3D-VRH mechanism at low temperatures. On the other hand, thin film samples exhibit a thickness dependent transition from 3D-VRH behaviour to Efros–Shklovskii variable range hopping behaviour.\cite{lu2014} These diverse reports suggest a role of sample preparation conditions and sample quality (defects, disorder etc) in deciding the nature of conduction at different temperature intervals. Although a change in conduction mechanism has been observed across the magnetic transition in all the cases, there is no apparent correlation between nature of conduction mechanism and the magnetic ordering.

As already mentioned, Ir-O-Ir bond angle plays a crucial role in deciding the magnetic and transport properties in this compound. The knowledge of temperature evolution of structural parameters and the bond angle is necessary to have a clear understanding of coupling between structure and physical properties. An earlier neutron diffraction\cite{crawford1994} and a synchrotron x-ray diffraction\cite{ranjbar2015} (xrd) studies have reported a decrease in this bond angle with the decrease in temperature. On the contrary, an xrd\citep{bhatti2014}  study has reported an increase in the bond angle with the reduction in temperature.

\begin{figure*}
%\hspace{-5mm}
\includegraphics[width=0.8\textwidth]{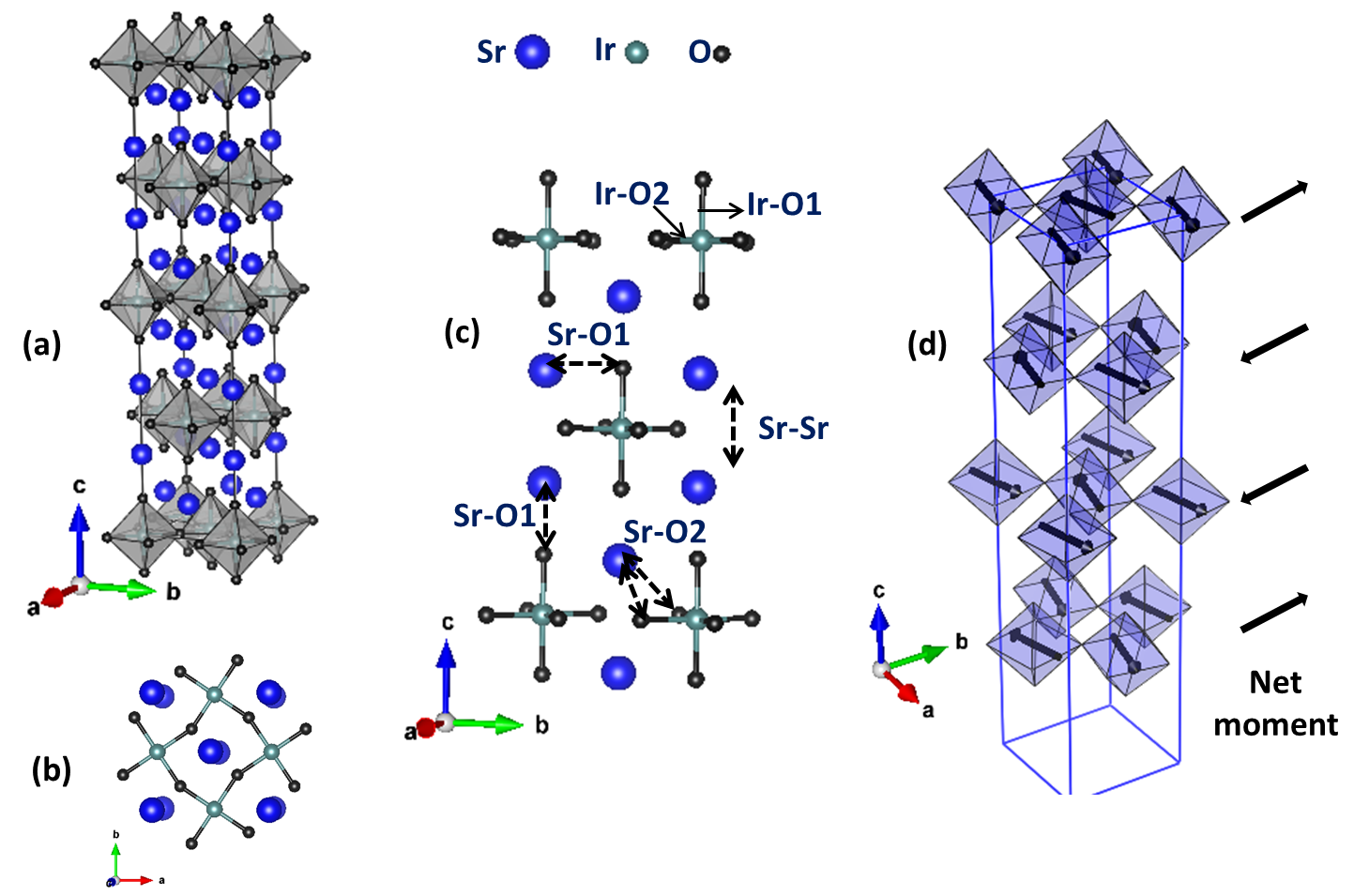}
%\vspace{-50mm}
\caption{The crystal and magnetic structure of Sr$_{2}$IrO$_{4}$.(a)Conventional unit cell of Sr$_{2}$IrO$_{4}$ (b)A part of unit cell in \emph{ab}-plane showing in-plane rotation of IrO$_6$ octahedra (c)A part of the unit cell in \emph{bc}-plane showing different bonds (d)Canted AFM structure of Sr$_{2}$IrO$_{4}$ as reported in the literature,\cite{ye2013} arrows on the right hand side indicate the direction of net magnetic moment from individual layers in the \emph{ab}-plane.}
\end{figure*}

The chemical substitution introduces disorder in the system which alters the magnetic and transport properties. Often, undoped parent compounds possess stoichiometric disorder. Recently, a suppression of metal-insulator transition has been reported in pyrochlore iridates (R$_2$Ir$_2$O$_7$) due to changes in R/Ir ratio.\cite{sleight2018} Although there are many reports on polycrystalline Sr$_2$IrO$_4$ with attempts of different chemical substitutions in search of superconductivity, there is a lack of studies focussing on the effect of disorder in this system. Apart from these behaviours, from the technological aspect, the system with compositional disorder has variation in its bond strengths. Hence, one can expect emergent phenomena at different length scales and hence exploiting the required properties as per our need. Towards this, we aim at studying the electronic, magnetic properties and its interplay in Sr$_{2}$IrO$_{4}$ compounds which are prepared under different preparation conditions.

Keeping all these above mentioned aspects in mind we have investigated the structural, magnetic, transport and magneto-transport properties of polycrystalline Sr$_2$IrO$_4$ and explore the effect of vacuum annealing on these properties. Annealed polycrytalline samples of Sr$_2$IrO$_4$ were found to be more insulating than the as-prepared samples in contrast to what has been observed in single crystal samples. The central findings of this work are : (a)Vacuum annealing reduces Sr/Ir ratio in addition to incorporating oxygen vacancy and enhances the tetragonality (c/a ratio) of the unit cell; (b) The annealed samples exhibit enhanced magnetization and resistivity at all temperatures; (c) connectivity between magnetic anisotropy and Ir-O-Ir (in-plane) bond angles; (d)The temperature dependence of transport behaviour is affected by annealing with implications of lesser disorder in annealed sample; (e)The magneto-transport studies have shown a significant enhancement in the negative magnetoresistance in the vacuum annealed sample. All these changes are associated with the changes in Ir-O-Ir in-plane bond angle and other structural parameters extracted from low temperature xrd measurements. The combined analysis of all the results suggests a disorder induced phase separated state at low temperatures in \emph{as-prepared} samples which gets suppressed after vacuum annealing.

\section{Experimental}
The polycrystalline samples of Sr$_{2}$IrO$_{4}$ were prepared by solid state reaction method using stoichiometric amounts of IrO$_2$ (99.9$\%$ pure) and SrCO$_3$(99.995$\%$ pure). A stoichimoetric mixture of the raw materials were ground for about 3 hours before calcining at 900$^\circ$C for 24 hours followed by 1000$^\circ$C for 24 hour and 1100$^\circ$C for 24 hours with intermediate grindings. The latter heat treatments were done in the pellet form. Heating and cooling rates utilized are 2$^\circ$C/minute and 1$^\circ$C/minute respectively. This method of preparation has been frequently employed in the literature for synthesizing polycrystalline Sr$_{2}$IrO$_{4}$ samples.\cite{klein2008,bhatti2014} To incorporate oxygen deficiency, the pellets of the samples were heated in an evacuated quartz tube ($\sim$10$^{-5}$mbar) at 600$^\circ$C for two different time intervals of 2.5 hours and 5 hours. The properties of the samples heated for 2.5 hours were similar to that of the as prepared compound. Hence we shall present a comparative study of the as prepared Sr$_{2}$IrO$_{4}$ (\emph{SIO-as prep}) and the sample which is vacuum annealed for 5 hours (\emph{SIO-anneal}).

The temperature dependent xrd patterns of all the samples were collected using Smart Lab 9kW rotating anode x-ray diffractometer with Cu-K$\alpha$ radiation. All the patterns were collected in the 2$\Theta$ range of 10$^{\circ}$ - 90$^{\circ}$ with the step size of 0.02$^{\circ}$. The scanning speed during the measurements was 1$^{\circ}$/minute. The low temperature measurements were performed inside a cryostat from RT to 12K for 15 different temperatures.

The energy dispersive x-ray analysis (EDS) measurements were performed on Nova Nano SEM-450 which is fitted with an energy dispersive x-ray spectrometer.

The temperature dependent dc magnetization measurements were performed using Magnetic Properties Measurement Systems (MPMS) in the temperature range of 300 K to 2 K at an applied magnetic field of 1T. The magnetic moment measuring system is from Quantum design, USA. The temperature dependent resistivity measurements were carried out using four probe method using commercial physical properties measurement system from Quantum design, USA. The magnetoresistance measurements were carried out at temperatures 300 K, 180 K, 150 K and 10 K, the magnetic field was applied up to 9 Tesla.

\section{Results and Discussions}
The crystal structure of Sr$_{2}$IrO$_{4}$ is shown in Fig.1(a).In Fig.2(a-b), we show the Reitveld refinement of the \emph{xrd} pattern of \emph{SIO-as prep} and \emph{SIO-anneal} samples collected at 300K. The experimental spectra shows a good agreement with tetragonal \emph{I4$_1$/acd} structural model, in line with the literature.\cite{crawford1994,klein2008,bhatti2014} The refined RT structural parameters along with the R-factors of refinement are shown in Table-I.  The RT xrd pattern of \emph{SIO-as prep} as well as \emph{SIO-anneal} also indicated the presence of small but identical amounts of SrIrO$_3$ impurity as shown in the inset of Fig.2(a). These impurities at 2$\theta$ $\sim$19.5$^\circ$ and $\sim$32.15$^\circ$ correspond to monoclinic  C2/c phase of SrIrO$_3$ compound.\cite{longo1971}
\begin{figure}
\hspace{-5mm}
\includegraphics[width=0.4\textwidth]{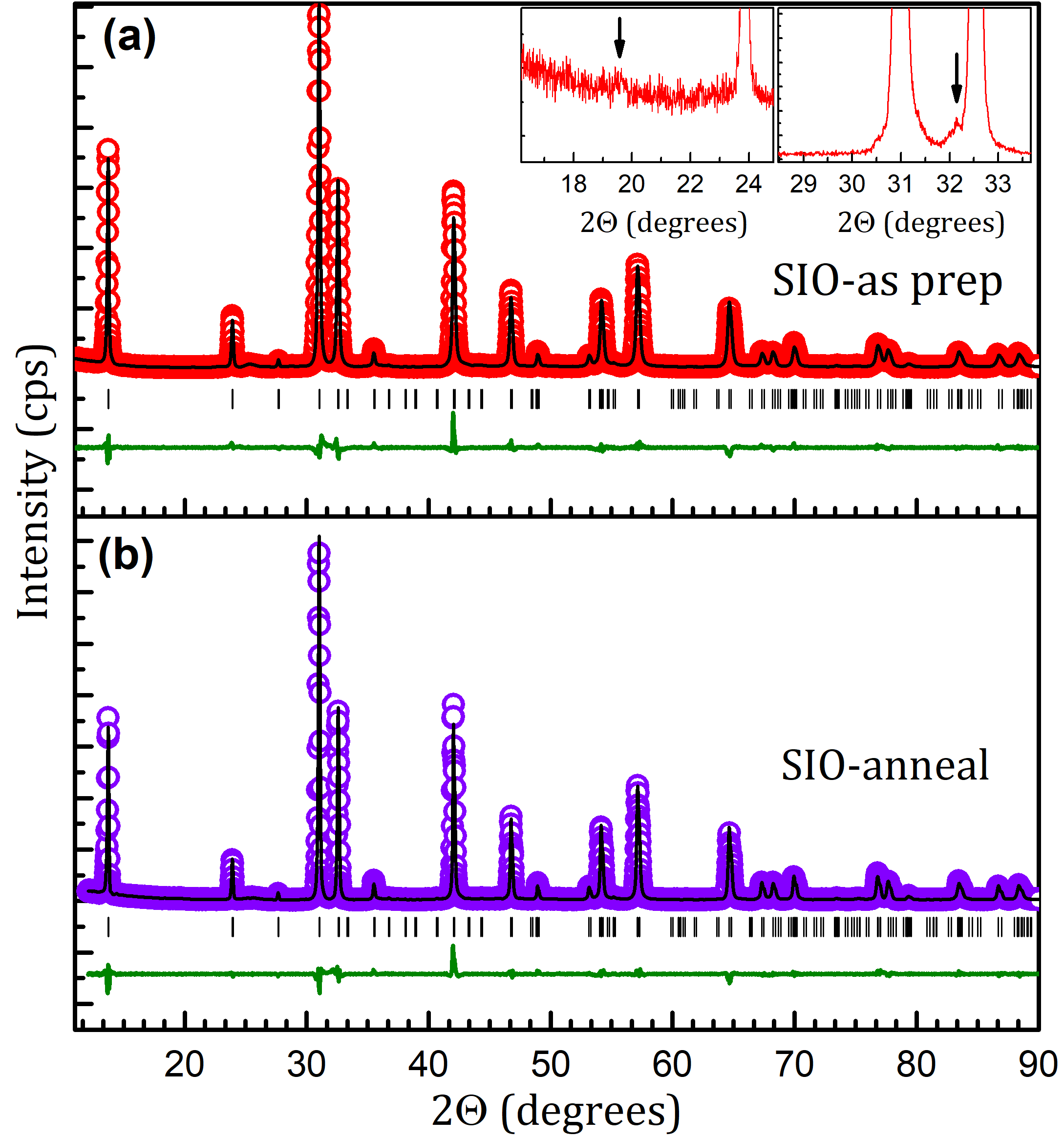}
%\vspace{-50mm}
\caption{Reitveld refinement of room temperature xrd pattern of (a)\emph{SIO-as prep} (b)\emph{SIO-anneal}. Experimental patterns are fitted to I4$_1$/acd tetragonal structure. The open and the line symbols correspond to the observed and calculated patterns, respectively. The Bragg reflections and the residual are given in black and green colours, respectively. The insets display enlarged view of the pattern with arrows indicating the positions of SrIrO$_3$ impurity phase.}
\end{figure}

The compositional analysis of the samples were carried out by Energy dispersive x-ray analysis (EDS) technique. The expected atomic percentages of Sr, Ir and O for a stoichiometric Sr$_{2}$IrO$_{4}$ are respectively, 28.57, 14.28 and 57.14. The atomic percentages of Sr, Ir and O obtained for \emph{SIO-as prep} are respectively $\sim$24.09, $\sim$11.61 and $\sim$64.3 while for \emph{SIO-anneal}, they are $\sim$25.21, $\sim$12.5 and $\sim$62.2.
The composition values indicate deficiency of Sr and Ir in both the samples. However, the ratio Sr/Ir is $\sim$2.08(9) in \emph{SIO-as prep} sample while it is $\sim$2.02(9) for \emph{SIO-anneal} sample. Although this ratio is similar within the error bar, there is a decreasing trend from \emph{SIO-as prep} to \emph{SIO-anneal} sample. Hence, \emph{SIO-anneal} has improved stoichiometry in comparison to \emph{SIO-as prep} sample. The atomic percentage values of oxygen are very much larger than the expected value in both the samples. This arises due to limited sensitivity of EDS to oxygen. The vacuum annealing process is expected to incorporate oxygen deficiency and the obtained atomic percentage values indicate a relatively lesser oxygen in \emph{SIO-anneal} sample. A possible reason for the observed off-stoichiometry in Ir may be due to the volatile nature of IrO$_2$ at high temperatures as reported in other Ir based oxides.\cite{sleight2018} The presence of Ir vacancy may result in the change of oxidation state of some Ir$^{+4}$ ions to non-magnetic Ir$^{+5}$. In essence, vacuum annealing has reduced the oxygen content along with a reduction of Sr/Ir ratio from $\sim$2.08 to $\sim$2.02.

\begin{table}
\caption{Crystallographic information for \emph{SIO-as prep} and \emph{SIO-anneal} samples at RT obtained from Rietveld refinement using \emph{I4$_1$/acd} space group. The atomic positions of Sr, Ir, O1 and O2 are (0,0.25,z),(0,0.25,0.375),(0,0.25,z) and (x, x+0.25,0.125)}

\resizebox{0.47\textwidth}{!}{%
\begin{tabular}{ c c c c c c c c c c c c c c c c c c  }
 \hline\hline \\

 & &\multicolumn{4}{c}{\emph{SIO-as prep}}  &&&  & & & \multicolumn{4}{c}{\emph{SIO-anneal} } & \\
 \cline{2-8} \cline{11-18} \\
% & U = 0eV & & 1eV& & 3eV& & 5eV & & &  0eV & &1eV&  &3eV& & 5eV  & \\
%\hline \\
Lattice parameters  &  & &  &   &  &       &      & &  & &   &   &    & &      &  &   \\
%\\
 a  &  & &  &   & 5.4975 \AA &       &      & &  & &   &   &    &5.4968 \AA &      &  &   \\
%\\
 c  &  & &  &   &25.791 \AA &  & & & & &  &  & &25.799 \AA & &   & \\
% \hline
\\
Atomic positions  &  & &  &   &  &       &      & &  & &   &   &    & &      &  &   \\
%\\
 Sr &  &z &  & &0.5506 &       &  &   &  &  &  &   & &0.5499 &       &   &   \\
%\\
 O1 &  &z &  & & 0.4692    &       &     & & & &  &    &  & 0.4697  &     &   &   \\
%\\
O2  &  &x &  & &0.1977 &       &    & & & &  &   &  &0.1944  &     &  &   \\
\\
%\hline
Refinement R factors&  & &  &   &  &       &      & &  & &   &   &    & &      &  &   \\
%\\
 R$_p$ &  & &  & & 9.42$\%$    &       &     & & & &  &    &  & 10.7$\%$  &     &   &   \\
 %\\
 R$_{wp}$ &  & &  & & 11.5$\%$    &       &     & & & &  &    &  & 13.5$\%$  &     &   &   \\
% \\
 R$_{exp}$ &  & &  & & 7.55$\%$    &       &     & & & &  &    &  & 7.39$\%$  &     &   &   \\
 \\
\hline\hline
\\
\end{tabular}}
\end{table}

Fig.3(a) shows the zero-field cooled(ZFC) and field-cooled(FC) magnetization curves for \emph{SIO-as prep} and \emph{SIO-anneal} samples at an applied magnetic field of 1T. Both the samples exhibit magnetic transition below around $\sim$238 K as seen from the inflexion point in the derivative of the ZFC magnetization curve, inset of Fig.3(a). The inverse susceptibility shown in Fig.3(c) do not obey Curie-Weiss behaviour in the entire paramagnetic region of both the samples, but in a limited temperature region. The transition temperature (T$_{N}$) obtained from the fitting are 233.7 K and 233.8 K, respectively for \emph{SIO-as prep} and \emph{SIO-anneal} samples. The estimated effective paramagnetic moment $\mu_{eff}$ value for \emph{SIO-as prep} is $\sim$0.58$\mu_B$ which is close to reported values\cite{bhatti2014} while $\mu_{eff}$ for SIO-anneal is $\sim$0.60$\mu_B$,higher than \emph{SIO-as prep}. The transition temperature marks the transition from paramagnetic to canted antiferromagnetic structure. The rise in the dc magnetization below T$_{N}$ suggests weak ferromagnetic behaviour arising due to the canted antiferromagnetic structure. The magnetic structure of Sr$_{2}$IrO$_{4}$ compound is shown in Fig.1(d). The J$_{eff}$=1/2 pseudospins are rotated in the \emph{ab}-plane by about 13$^\circ$ and are locked to in-plane IrO$_6$ octahedral rotation due to strong spin orbit coupling.\cite{ye2013} This canting of spins results in a net magnetic moment along \emph{b}-direction in each Ir-O layer which are arranged in ($\uparrow\downarrow\downarrow\uparrow$)manner along \emph{c}-direction as shown in Fig.1(d).\cite{ye2013,kim2009} The intralayer magnetic interaction which is dominated by Heisenberg exchange coupling ($\sim$60meV) is stronger  compared to interlayer interaction.\cite{haskel2020} Both these interactions (intra and interlayer) are proposed to be necessary to stabilize the observed canted AFM magnetic structure.\cite{katukuri2014} The canting of moments in the \emph{ab}-plane arises due to DM-interaction which is responsible for weak ferromagnetic behaviour of the compound.

Apart from this, the FC and ZFC magentization curves show bifurcation below $\sim$140K (T$_{irr}$) in both the samples. The ZFC curve exhibits a dip below $\sim$65K in both the samples. It can also be seen that the difference between FC and ZFC curves is higher for \emph{SIO-as prep} below $\sim$40K, Fig.3(b). The observation of the bifurcation in the FC and ZFC magentization curves suggest magnetic anisotropy\cite{joy1998} existing in these systems. Based on the extent of bifurcation of the FC and ZFC magnetizations one can conclude that the magnetic anisotropy is more in the case of \emph{SIO-as prep} as compared to the \emph{SIO-anneal} sample.

In the system that is magnetically isotropic, in the presence of the external magnetic field, the system shows no directional dependence in terms of the alignment of the spins while in the case of the system that exhibit magnetic anisotropy, there occurs preferential alignment of the spin along a particular direction in response to the external field. Magnetic anisotropy is usually observed in the systems that display properties like spin glass, cluster glass, superparamagnetism, have strong spin orbit coupling or any short range magnetic ordering etc. When the sample is cooled in zero field, the spins are randomly oriented but cooling the sample in the presence of field leads to the tendency of the spins to align in the direction of the applied field. This tendency depends on the magnetic anisotropy of the system and the strength of the applied field. If the system shows more anisotropy, higher fields are required to align the spins in the direction of the applied field and hence to higher magnetisation.

In the compound under study, reports show that the magnetic anisotropy is driven mainly by (a) strong spin orbit coupling, below T$_{N}$ which leads to canted antiferromagnetic structure and (b) possibility of stabilisation of frustrating magnetic order\cite{chikara2009} and co-existence of cluster glass phase along with canted AFM phase\cite{manna2014} at low temperatures.

\begin{figure}
\hspace{-5mm}
\includegraphics[width=0.5\textwidth]{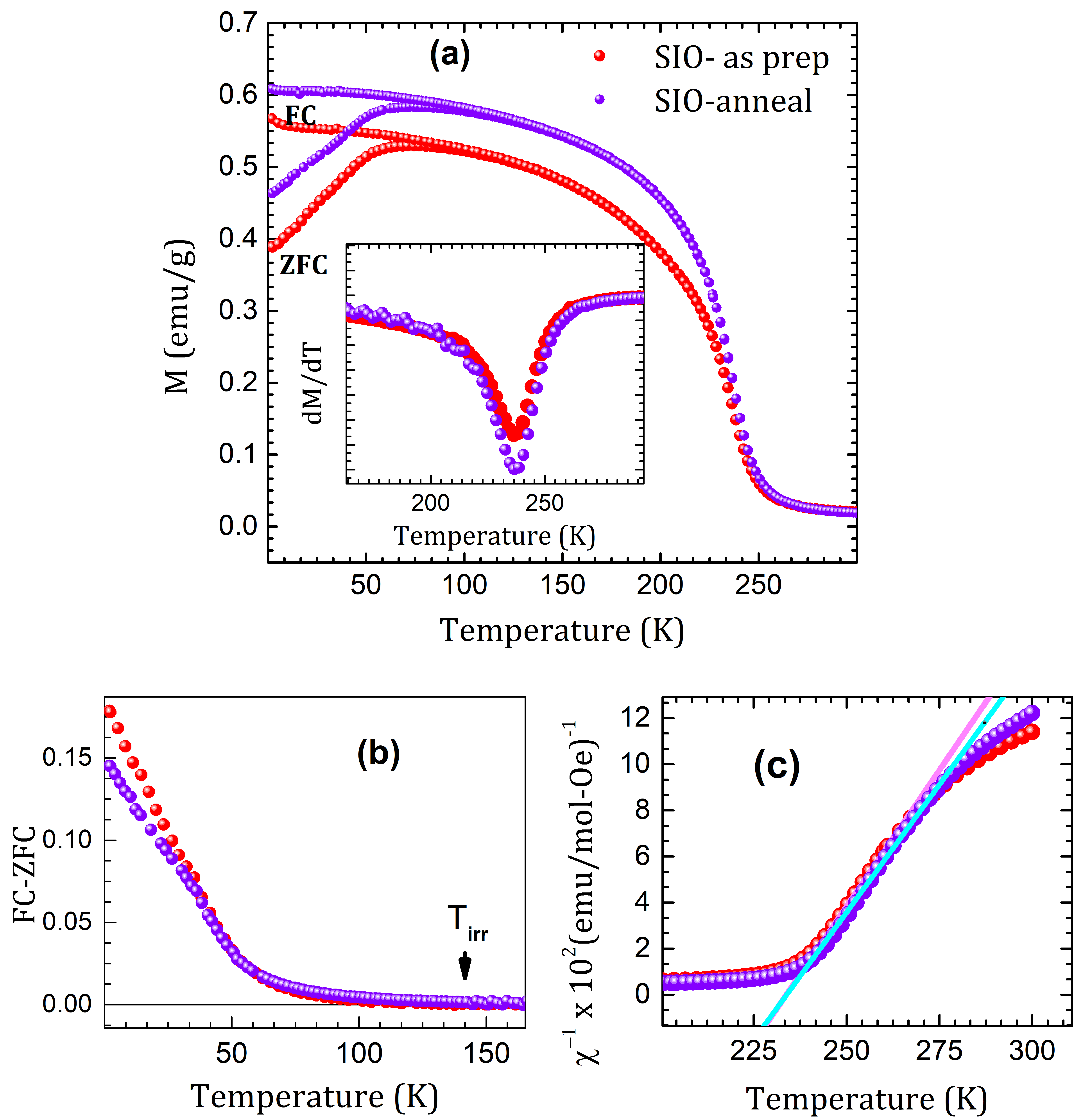}
%\vspace{-50mm}
\caption{(a)ZFC and FC magnetization curves for SIO-parent and SIO-anneal samples at 10kOe field, inset shows the derivative of ZFC magnetization curve for both the samples; (b)Difference between FC and ZFC for both the samples; (c) Inverse susceptibility vs temperature curves, solid lines indicate Curie-Weiss fit to the experimental data}
\end{figure}

The stabilisation of spin orbit coupling in the compounds under study is expected to manifest in the structural parameters. Towards this, we now investigate the results of the temperature dependent x-ray diffraction experiments collected on both the samples. Figs.4(a-d) shows the temperature variation of lattice parameters, unit cell volume and c/a ratio of \emph{SIO-as prep} and \emph{SIO-anneal} samples, extracted from the Rietveld refinement of temperature dependent xrd measurements. The a-parameter shows a decreasing trend while the c-parameter increases with the decrease in temperature, in agreement with the literature. The a-parameter decreases monotonically until around 80K with a variation rate of $\Delta$a/a $\sim$0.24$\%$ in both the samples. While, below 80K, this rate of variation is reduced significantly ($\Delta$a/a $\sim$0.02$\%$). Similarly, the c-parameter also decreases from RT to around 270K in both the samples. Around the region of T$_{N}$, we observe a change in slope of a-parameter and a small hump in c-parameter, insets of Fig.4(a) and Fig.4(b), respectively. Below 250K to around 60K, c-parameter increases with a variation rate of $\sim$0.03$\%$ in both the samples. Below 60K, rate of variation decreases and the c-parameter of \emph{SIO-as prep} sample increases by about 0.008$\%$ while that of \emph{SIO-anneal} by 0.005$\%$.
The unit cell volume decreases while the c/a ratio which is a measure of tetragonality of the system increases with the decrease in temperature. The a-parameter is lower in \emph{SIO-anneal} sample compared to \emph{SIO-as prep} whereas the c-parameter is lower in \emph{SIO-as prep} at all temperatures. The effect of annealing is the preservation of the unit cell volume, while the tetragonality c/a increases when we go from \emph{SIO-as prep} to \emph{SIO-anneal} sample at all temperatures, Fig.4(c and d). The anisotropic temperature behaviour of lattice parameters can be explained on the basis of temperature evolution of relevant bond distances and bond angles.

\begin{figure}
\hspace{-5mm}
\includegraphics[width=0.5\textwidth]{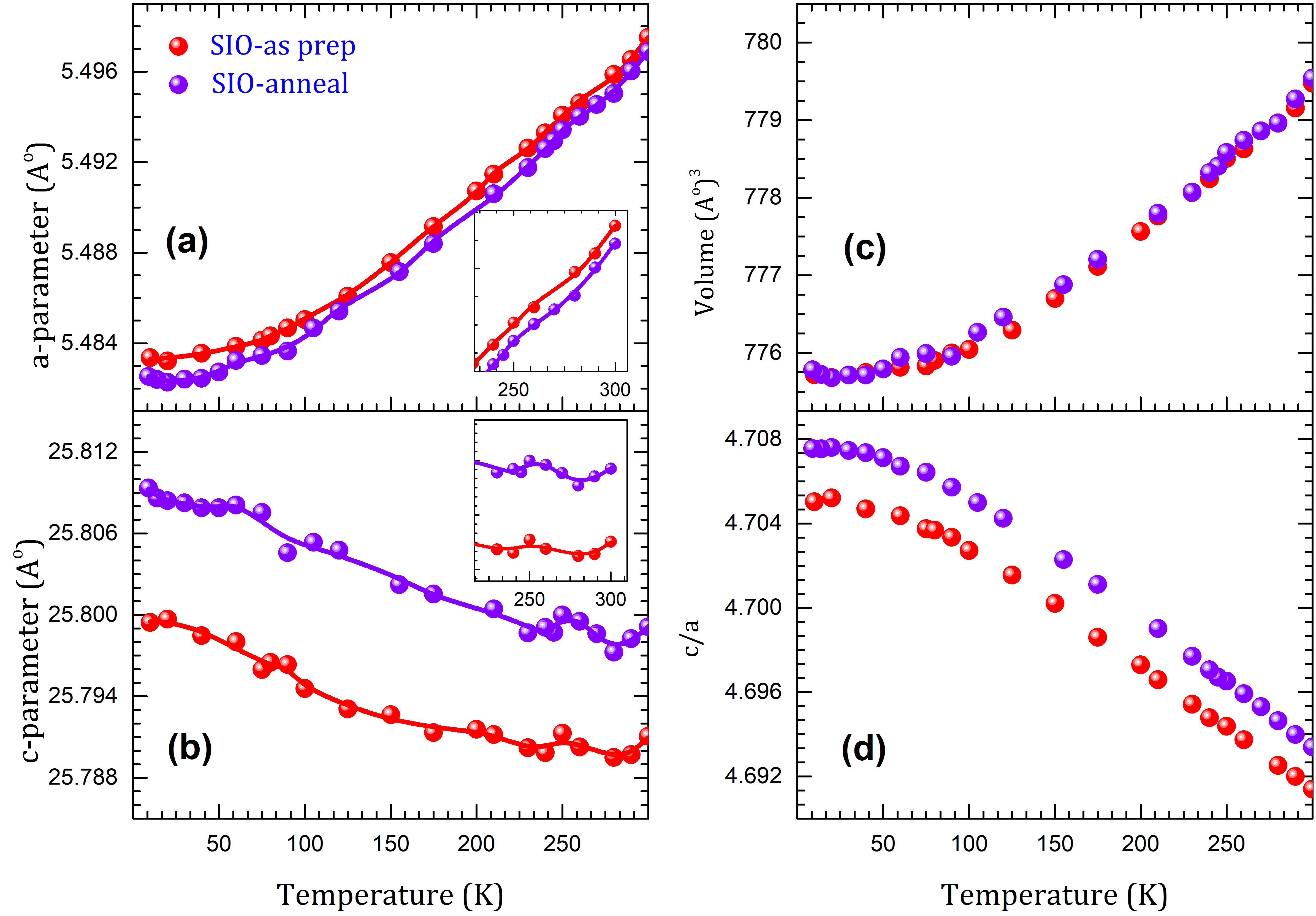}
%\vspace{-50mm}
\caption{(a-d) Temperature evolution of lattice parameters, unit cell volume and c/a ratio of \emph{SIO-as prep} and \emph{SIO-anneal} samples obtained by Reitveld refinement. The error bar is within the symbol size. The line represents guide to the eye. }
\end{figure}

\begin{figure*}
\hspace{-5mm}
\includegraphics[width=0.8\textwidth]{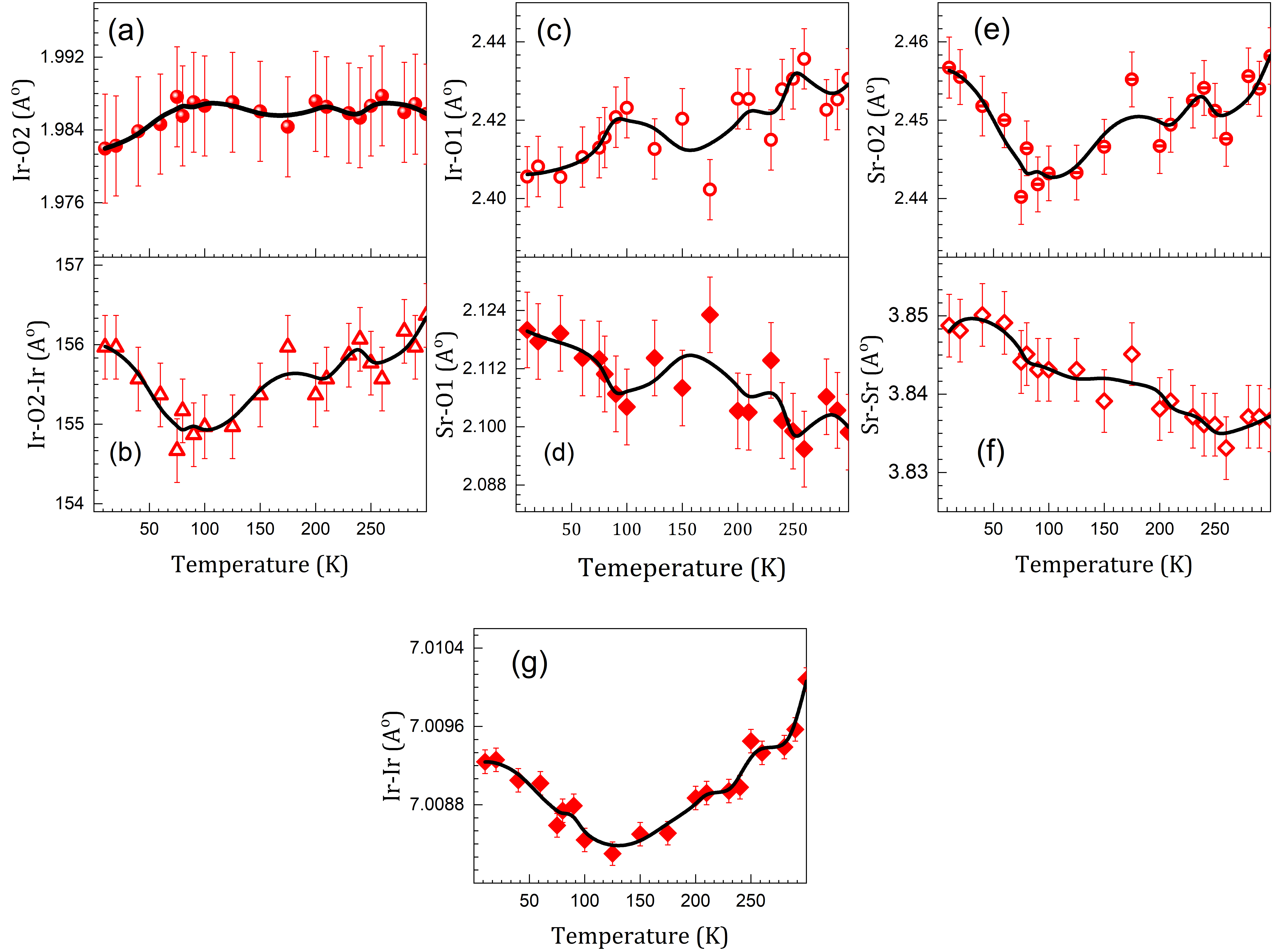}
%\vspace{-50mm}
\caption{Temperature evolution of bond lengths and bond angle in \emph{SIO-as prep} sample (a) Ir-O2 basal bond length; (b) Ir-O2-Ir in-plane bond angle; (c) Ir-O1 apical bond length; (d) Sr-O1 bond length parallel to \emph{c}-axis; (e) Sr-O2 shorter bond length; (f) Sr-Sr bond length parallel to \emph{c}-axis; (g) Ir-Ir out of plane bond distance.The line represents guide to the eye.}
\end{figure*}

\begin{figure*}
\hspace{-5mm}
\includegraphics[width=0.8\textwidth]{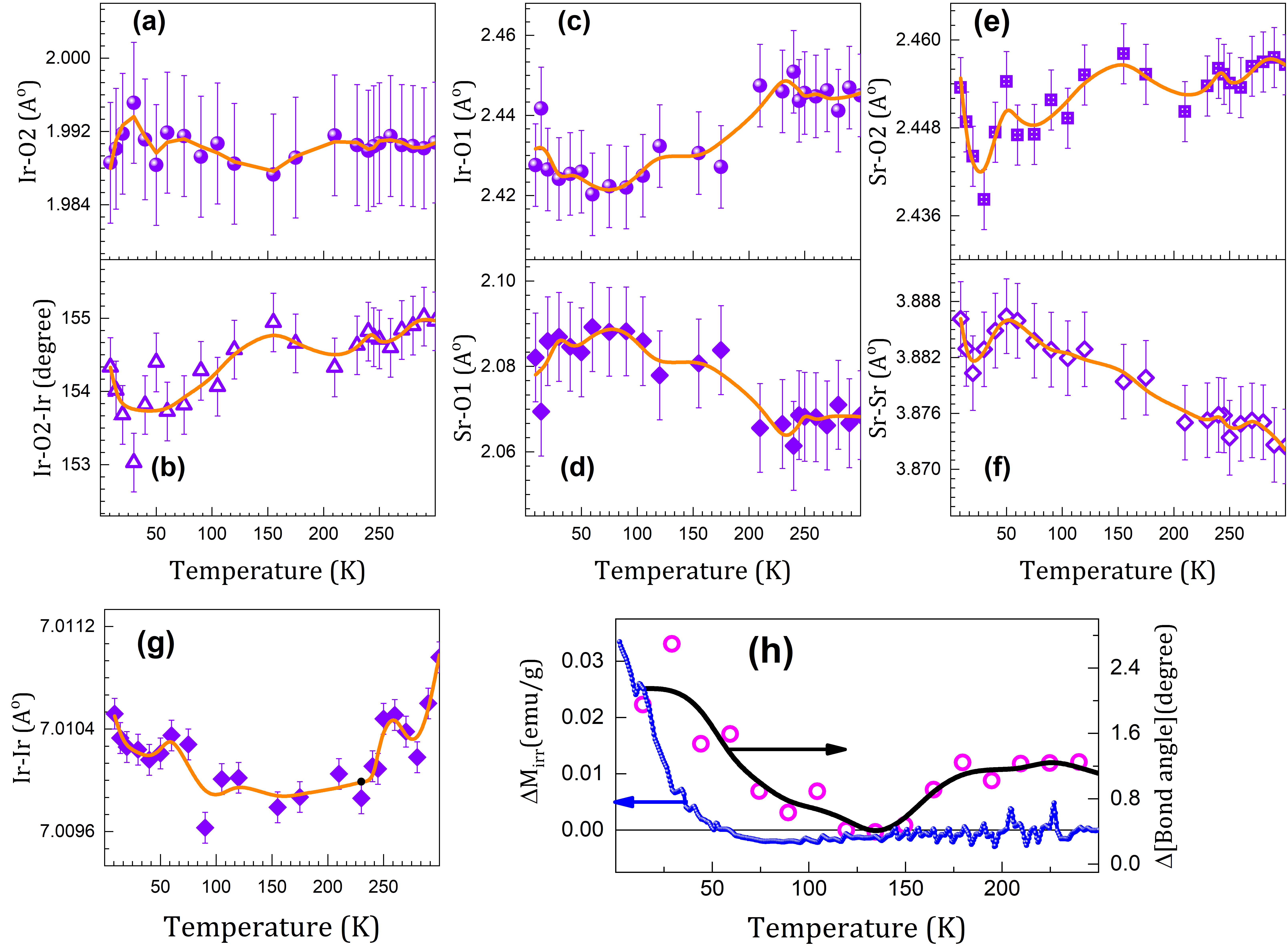}
%\vspace{-50mm}
\caption{Temperature evolution of bond lengths and bond angle in \emph{SIO- anneal} sample (a) Ir-O2 basal bond length; (b) Ir-O2-Ir in-plane bond angle; (c) Ir-O1 apical bond length; (d)Sr-O1 bond length parallel to \emph{c}-axis; (e) Sr-O2 shorter bond length; (f) Sr-Sr bond length parallel to \emph{c}-axis; (g) Ir-Ir out of plane bond distance;(h) Comparison of \textbf{the difference in the }irreversible magnetization between both the samples and the difference in the in-plane Ir-O2-Ir bond angles between both the samples.The line represents guide to the eye.}
\end{figure*}

There are two kinds of Ir-O, Sr-O1 and Sr-O2 bonds as shown in Fig.1(c). The temperature evolutions of some of these bonds are shown in Fig. 5(a-g) for \emph{SIO-as prep} and Fig.6(a-g) for \emph{SIO-anneal} samples. The Ir-O2 in-plane bond length shows negligible variation from RT to around 80K for \emph{SIO-as prep} sample, Fig.5(a). At the same time, there is a overall decrease in  Ir-O2-Ir in-plane bond angle from RT to $\sim$80K, Fig.5(b). Below 80K, Ir-O2 bond length shows a decreasing trend while Ir-O2-Ir bond angle increases.
Hence, from RT to $\sim$80K, the decrease in a-parameter is dominated by decrease in Ir-O2-Ir bond angle while below 80K, the increase in Ir-O2-Ir bond angle along with the decrease in Ir-O2 results in lesser variation of a-parameter.

In general, lattice parameters are expected to decrease with the decrease in temperature due to reduction in lattice vibrations. The negative thermal expansion in the c-parameter has been attributed to the in-plane rotations of the IrO$_{6}$ octahedra as allowed by the space group \emph{I4$_1$/acd}. At the local bond length level, such increment in the c-parameter can be understood as follows. The Ir-O1 and Sr-O1 bonds shown in Fig.5(c-d), which are parallel to c-axis exhibit opposite temperature dependence and hence do not contribute to increase in \emph{c}-parameter. On the other hand, Sr-Sr bonds increase with the decrease in temperature as shown in Fig.5(f). This increase in Sr-Sr distance is related to the Ir-O2-Ir in-plane bond angle variation with temperature. The Ir-O2-Ir bond angle changes due to change in the position of O2 in the ab-plane. The temperature variations of Sr-O2 bond lengths follow the Ir-O2-Ir bond angle variation. The bond angle variation causes the shorter Sr-O2 bond length to decrease with temperature from RT to $\sim$80K, Fig.5(e). Hence, it is favourable for the Sr ions parallel to \emph{c}-axis to move away from each other for minimizing repulsion with basal oxygen ion cloud. This leads to the unusual expansion of the c-axis with the reduction in temperature. Below 80K, the increase in Ir-O2-Ir bond angle increases Sr-O2 shorter bond which further leads to lesser variation in Sr-Sr bond length. Hence c-parameter variation is suppressed at low temperatures.

Similar to \emph{SIO-as prep} sample, Ir-O2 bond length and Ir-O2-Ir bond angle decide temperature dependence of \emph{a}-parameter in \emph{SIO-anneal} sample, Fig.6(a-b). The increase in c-parameter with temperature can be explained in a similar manner as that of \emph{SIO-as prep} sample. The Ir-O1 and Sr-O1 bonds which are parallel to z-axis show opposite temperature behaviour, Fig.6(c-d). The Sr-O2 shorter bond length shows a decreasing trend which leads to the increase of Sr-Sr bond length, Fig.6(e-f).

In \emph{SIO-anneal} sample, we have seen that c/a ratio increases at all temperatures compared \emph{SIO-as prep}. When we compare the lattice parameters of both the samples, we observed that the a-parameter is larger in the case of the \emph{SIO-as prep} as compared to the \emph{SIO-anneal} sample while for the case of the c-parameter, the behaviour is opposite. Such behaviour could be due to the changes in the composition of the sample. From the compositional analysis, the \emph{SIO-as prep} has Ir vacancies due to which Ir$^{+5}$ ions can be present in the sample. Annealing process has improved the stoichiometry thereby converting some of Ir$^{+5}$ to Ir$^{+4}$. The ionic radius of Ir$^{+4}$ is higher than that of Ir$^{+5}$ which results in the increase of Ir-O1 and Ir-O2 bond distances in \emph{SIO-anneal} sample. The decrease in a-parameter of \emph{SIO-anneal} is due to the decrease in Ir-O2-Ir bond angle which dominates the increase in Ir-O2 bond length. The increase in Ir-O1 bond length contributes to increase in \emph{c}-parameter of \emph{SIO-anneal} sample.

Now we shall explain the consequences of structural parameter variation on the magnetic properties of \emph{SIO-as prep} and \emph{SIO-anneal} samples. The magnetism in Sr$_2$IrO$_4$ is stabilized by a stronger intralayer and weaker interlayer exchange interactions leading to a planar canted AFM structure with J$_{eff}$-1/2 moments in the ab-plane and ($\uparrow\downarrow\downarrow\uparrow$) alignment of net moments along the c-direction, as shown in Fig.1(d).
The strong spin orbit coupling, locks the moments to the IrO$_6$ octahedral rotation. Hence, the decrease in Ir-O2-Ir bond angle with temperature suggests an increase in canting angle until $\sim$125K in \emph{SIO-as prep} sample. Such a decrease in Ir-O2-Ir bond angle in Sr$_2$IrO$_4$ is observed in previous neutron diffraction measurements.\cite{crawford1994} This decrease in bond angle is accompanied by decrease in Ir-Ir out of plane bond length until $\sim$125K as shown in Fig.5(g), strengthening the interlayer coupling. The decrement of both the Ir-O2-Ir bond angle and the Ir-Ir bond distances suggests the setting up of the intralayer and interlayer interactions and hence the setting up of canted AFM structure\cite{manna2014}. In the temperature range 125 to 80K, Ir-O2-Ir bond angles remain almost the same suggesting a competing interaction at play. Below $\sim$125 K, the Ir-Ir bond lengths exhibit an increment down to low temperatures. The increase in Ir-O2-Ir bond angle suggests a decrease in canting and increase in Ir-Ir out of plane distance further weakens interlayer exchange interaction. These results indicate a deviation from canted AFM structure in \emph{SIO-as prep} sample below $\sim$80K.

The Ir-O2-Ir bond angle for \emph{SIO-anneal} samples are less than that for \emph{SIO-as prep} sample at all temperatures, Fig.6(b). This indicates an increased canting and consistent with the observation of higher magnetization for \emph{SIO-anneal} compared to \emph{SIO-as prep} sample.  With the decrease in temperature, the Ir-O2-Ir bond angle in \emph{SIO-anneal} shows an increment $\sim$150K and then a decrement upto $\sim$75K. In the temperature range 75 to 40 K, the Ir-O2-Ir bond angles remain almost the same. This suggests a competing interaction that opposes the stabilization of canted AFM structure. Below $\sim$30K , a small increasing trend is observed. There is an overall decrease in the bond angle indicating a dominant canted AFM phase until $\sim$75K. The Ir-Ir out of plane bond length decreases from RT to around 230K with a pronounced feature $\sim$240K, Fig.6(g). From 230K to $\sim$90K this bond length shows a negligible variation. Below 90K, an increasing trend is observed. The magnitude of increase in this bond length below 90K to low temperature is lesser compared to that observed in \emph{SIO-as prep} sample. All these results suggest a dominant canted AFM magnetic structure in \emph{SIO-anneal} sample until $\sim$75K, below which magnetic structure may deviate as reported in the literature\cite{manna2014, chikara2009} with the possibility of stabilisation of disordered magnetic phase along with the canted antiferromagnetic structure. It is interesting to note that below T$_{irr}$, the behaviour of the difference in the irreversible magnetisation between both the samples is in line with the difference in the Ir-O2-Ir bond angle between both the samples, Fig.6(h).The structural link with the magnetism is summarised pictorially in Fig.7. From this figure, it is clear that with vacuum \emph{annealing}, the range of stabilisation of canted AFM structure is enhanced by 50 K as compared to the \emph{as prepared} sample and hence the strength of the canted AFM structure. From Fig.7, it is interesting to note that in $\emph{SIO-as prep}$ and $\emph{SIO-anneal}$ samples, $\sim$ 175 K and 150 K, respectively a slight increment in the Ir-O2-Ir bond angles is observed. Such an increment suggests the role of hybridisation between the Ir 5$\emph{d}$ and O 2$\emph{p}$ orbitals in the setting up of the intralayer exchange interaction.

\begin{figure}
\hspace{-5mm}
\includegraphics[width=0.5\textwidth]{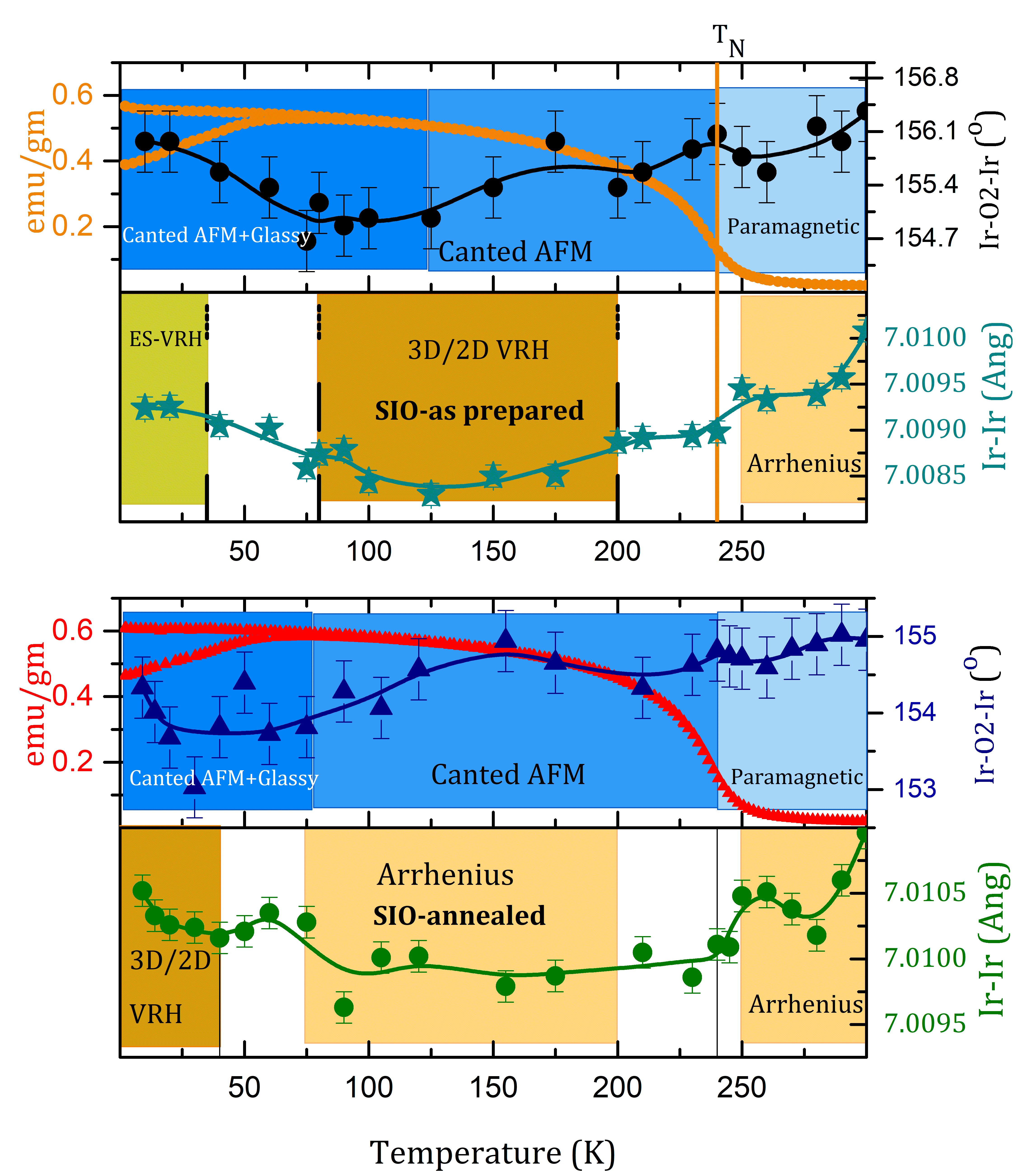}
%\vspace{-25mm}
\caption{Pictorial representation of the results of the transport,magnetic properties,Ir-O2-Ir bond angle and Ir-Ir out of plane bond distances as a function of temperature for \emph{SIO-as prep} (upper panel) and \emph{SIO-anneal} (lower panel) samples.The line represents guide to the eye.}
\end{figure}

We now investigate the link between the magnetic and structural properties on to the transport properties. Fig.8 shows the resistivity variation of \emph{SIO-as prep} and \emph{SIO-anneal} samples with temperature. Both the samples exhibit semiconducting like behaviour in the entire temperature range. The resistivity of \emph{SIO-anneal} sample is higher than that of \emph{SIO-as prep} at all the temperatures. It can be seen that resistivity rises very rapidly below around 100K upto the lowest temperature studied.

\begin{figure}
\hspace{-5mm}
\includegraphics[width=2.5in]{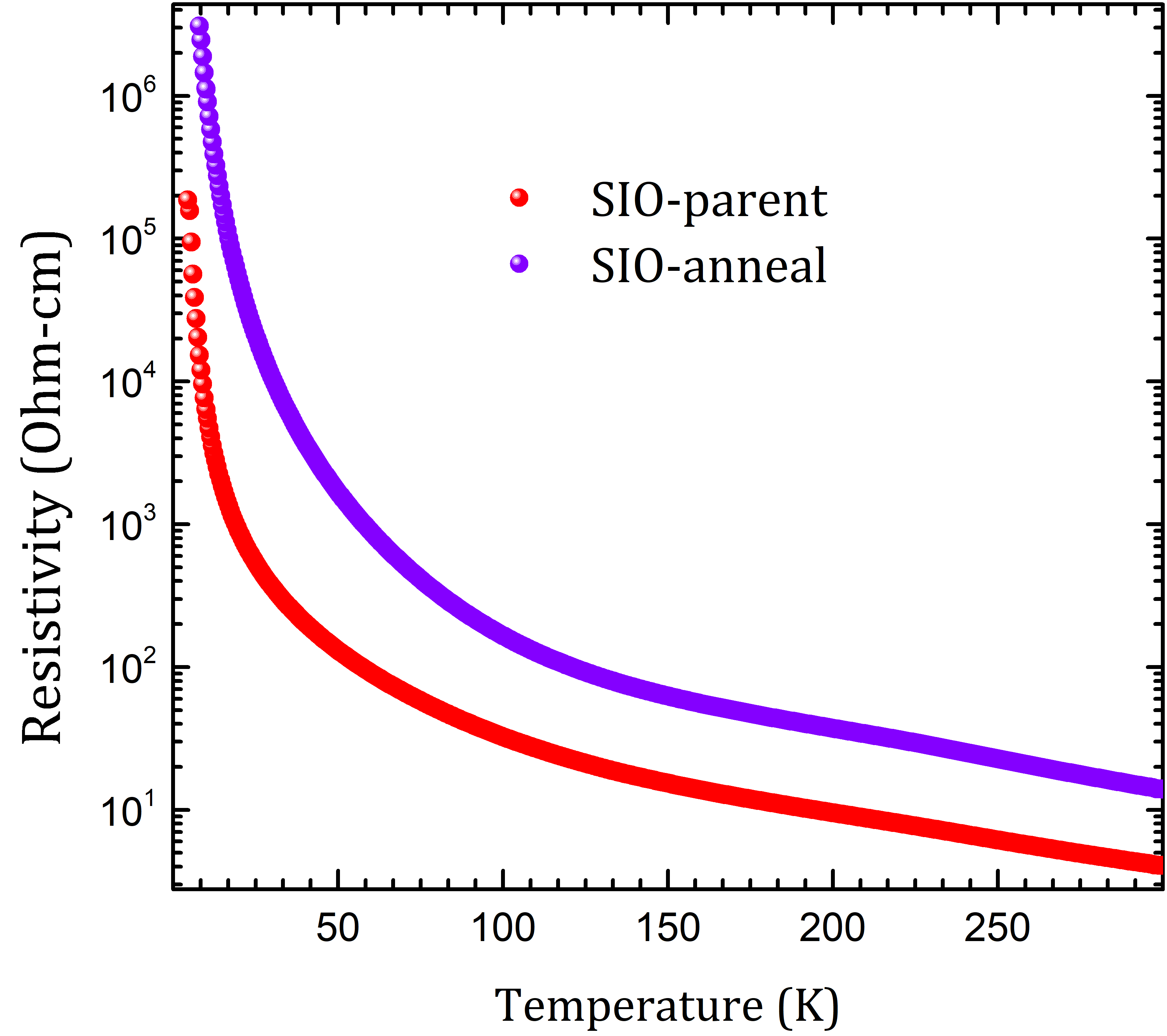}
%\vspace{-50mm}
\caption{Resistivity variation with temperature for \emph{SIO-as prep} and \emph{SIO-anneal} samples. }
\end{figure}

\begin{figure*}
\hspace{-5mm}
\includegraphics[width=5in]{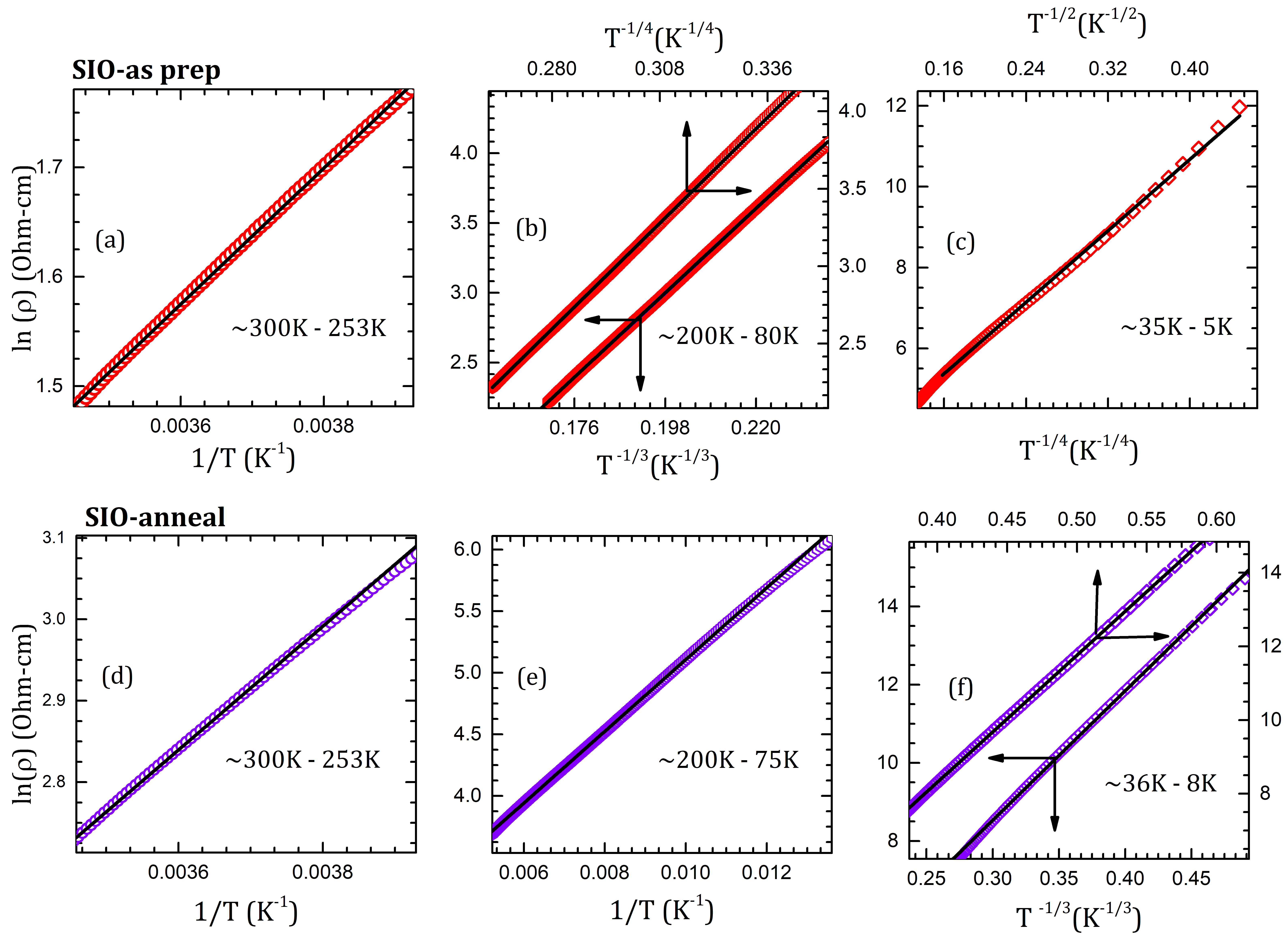}
%\vspace{-50mm}
\caption{Upper panel-resistivity variation of \emph{SIO-as prep} showing (a)Arrhenius behaviour in the range 300K to 253K (b) Variable range hopping from $\sim$200K to 80K (c)Efros–Shklovskii (ES) variable-range hopping at low temperature. Lower panel-resistivity fitting for \emph{SIO-anneal} sample: Arrhenius equation fitting in (a) from 300K to 253K (b) from 200K to 75K and (c)Variable range hopping at low temperatures. Solid line indicates the fit and the symbols are experimental data. }
\end{figure*}

It is possible to identify three different temperature regions having a distinct resistivity behaviour in both samples as shown in Fig.9(a-c) for \emph{SIO-as prep} sample and in Fig.9(d-f)in \emph{SIO-anneal} sample. In the paramagnetic phase of both the samples, resistivity can be modeled by Arrhenius equation given by $\rho=\rho_0\ \exp(E_a/k_B T)$, where $E_a$ is the activation energy, $k_B$ is Boltzmann constant and $\rho_0$ is a constant. In Arrhenius hopping model, the charge carriers undergo thermal excitations of charge carriers from localized state to the extended states. A plot of $ln\rho$ vs 1/T is shown in Fig.9(a) and Fig.9(d) for \emph{SIO-as prep} and \emph{SIO-anneal} samples respectively, which are straight lines in the temperature range of 300K to $\sim$253K. The activation energy obtained from the fitting are 107 meV and 130 meV for \emph{SIO-as prep} and \emph{SIO-anneal} samples, respectively. These values are closer to the reported values for single crystal samples\cite{ge2011} and also to those obtained from optical absorption measurements.\cite{kim2008}

Below the magnetic transition temperature, from 200K to $\sim$80K, \emph{SIO-as prep} obeys Mott's variable range hopping (VRH) equation given by,\cite{mott1968}
\begin{center}
 $\rho(T)=\rho_0\ \exp(T_M/T)^{1/p+1}$
 \end{center}
 where $\rho_0$ is a constant, \emph{p} is the dimensionality of the system and $T_M$ is a characteristic temperature which is related to electron localization length $\xi$. VRH model is applicable in temperature region where there is slowly varying DOS around the region of the Fermi level and the electrons at the Fermi level are localized. In this model, conduction occurs through hopping of the charge carriers from one localized state to another localized state. In 3-dimensions (\emph{p=3}, the relation between the localization length $\xi$ and T$_M$ is given by the equation,\cite{rosenbaum1991}
 \begin{center}
 $ T_M=18/[k_B\xi^3N(E_f)]$
 \end{center}
 Here, $N(E_f)$ is the number of available electron states per unit volume at the fermi level, $E_f$.
In 2-dimension (\emph{p=2}), it is given by\cite{pepper1974}

\begin{center}
$ T_M=27/[k_B\xi^2N(E_f)]$
\end{center}
Here, $N(E_f)$ is the number of available energy states per unit area at the fermi level, $E_f$.

The plots of both ln$\rho$ with $T^{-1/3}$ and with $T^{-1/4}$ are straight lines as shown in the Fig.8(b) in almost the same temperature range from $\sim$200K to $\sim$80K. We have obtained a value of $\rho_0$=0.01456 $\Omega$-cm and 0.08713 $\Omega$-cm respectively for 3D and 2D VRH fitting. Also, the obtained values of $T_M$ for 2D and 3D cases are respectively  $3.67 \times 10^5 K$ and $20.737 \times 10^3 K$.
Lu \emph{et al}\cite{lu2014} have reported similar values of 3D-VRH fitting parameters for epitaxial thin film samples in the higher thickness limit while Bhatti \emph{et al}\citep{bhatti2014} have obtained similar 2D-VRH parameters in the bulk sample and in the intermediate temperature range.

Our results show that we could fit the resistivity curves in the temperature range 200 to 80 K using both 2D and 3D VRH models. It is important to note that in this compound where SOC plays important role, there is a strong link between the transport and the magnetic properties. The canted AFM structure is known to exhibit weak ferromagnetic behaviour. With decrease in temperature, the rise in the dc magnetization around 240 K is due to the weak ferromagnetism and further decrement in the temperature leads to the setting up of the antiferromagnetic interaction between the layers. Hence, data fitted with both 2D and 3D VRH mechanism suggest the setting up of the magnetic interactions both within and between the layers.It is important to note that the interaction between the layers is a weak one.

Using the co-efficient of electronic part of specific heat $\gamma$, the estimated density of states per unit volume at $E_f$ is $\sim$ 6 $\times$ 10$^{28}$/eV m$^3$.\citep{kini2006, bhatti2014} The electron localization length $\xi$ obtained using $T_M$ and $N(E_f)$ for 3D case  is $\sim$2.12{\AA}, which is very close to in-plane Ir-O bond length. It is important to note that the value of $\xi$ may deviate due to deviation in $N(E_f)$ from the actual value in the temperature domain of VRH fitting.
The ratio of mean hopping distance $R_M$ to electron localization length $xi$ given by,
\begin{center}
$R_M/\xi=(3/8)(T_M/T)^{1/4}$
\end{center}
 is always greater than 1 in the temperature range 200K to 80K, which supports the applicability of Mott's VRH equation.\cite{rosenbaum1991} We could not calculate the corresponding localization length for the 2D case due to lack of information regarding $N(E_f)$.
 \\

The cross over from Arrhenius behaviour to variable range hopping mechanism of resistivity is generally observed in disordered semiconductors with VRH occurring at low temperatures.\cite{mott2012} As we have seen in EDS analysis, \emph{SIO-as prep} has stoichiometric disorder in the form of Ir deficiency. This may be the reason behind the occurrence of VRH behaviour at a relatively higher temperature. On the other hand, in almost the same temperature interval, \emph{SIO-anneal} sample follows Arrhenius equation similar to that in the paramagnetic phase but with a lesser activation energy of 50 meV, Fig.9(e). The observation of Arrhenius behaviour in the intermediate temperature range is previously reported by Kini \emph{et al.}\cite{kini2006} in polycrystalline Sr$_2$IrO$_4$ prepared by solid state reaction method starting from Ir metal and with a relatively lower temperature of heat treatment compared to the present study. It is to be noted that, EDS analysis indicated a change in the $Sr/Ir$ ratio from $\sim$2.08 for \emph{SIO-as prep} to $\sim$2.02 for \emph{SIO-anneal}. Hence, \emph{SIO-anneal} is less disordered than \emph{SIO-parent} leading to Arrhenius behaviour in the intermediate temperature range.

At low temperatures, below $\sim$35K, the resistivity of \emph{SIO-as prep} can be described with the Efros–Shklovskii (ES) variable-range hopping equation given by,\cite{efros1975}
\begin{center}
 $\rho(T)=\rho_0\ \exp(T_{ES}/T)^{1/2}$
\end{center}
 where, T$_{ES}$ is a characteristic temperature. A plot of ln$\rho$ vs $T^{-1/2}$ is shown in Fig.9(c) which fits fairly well with a straight line. In the ES VRH model, Coulomb gap appears in the region around the Fermi level. This arises due to the introduction of long-range Coulomb interaction between the electrons in the localized states.\cite{efros1975} A value of T$_{ES}$=487.52K is obtained from the fitting. The ratio of average hopping distance $R_{ES}$ to the localization length $\xi$ must be greater than 1 in the temperature range of fitting for the validity of ES-VRH hopping. This ratio is given by,

 \begin{center}
 $R_{ES}/\xi=(1/4)(T_{ES}/T)^{1/2}$
 \end{center}

 Using the value of $T_{ES}$, we found this ratio to be greater than 1 upto $T\leq30K$. Thus, in \emph{SIO-as prep} sample ES-VRH picture is valid from the lowest temperature to 30K. The expression for the Coulomb gap $\Delta_{CG}$ is given by,

 \begin{center}
 $\Delta_{CG}\simeq k_B (T^3_{ES}/T_M)^{1/2}$
 \end{center}
 The value of this gap is found to be $\sim$1.5meV.
  Recently, a similar temperature dependence and a cross over from Mott 3D-VRH to ES-VRH mechanism at low temperatures  has been reported in thin film samples of Sr$_2$IrO$_4$\cite{lu2014} and also in La-doped single crystal samples.\cite{chen2015}In literature, T$^{-1/2}$ behaviour is also observed in granular metal systems consisting of metallic grains embedded in an insulating media.\cite{sheng1973} A similar phase separation has been observed in La-doped single crystal samples probed by scanning tunnelling microscopy along with magneto-transport and neutron diffraction measurements.\cite{chen2015}Thus the observation of ES-VRH hopping may also be due to inhomogeneity at the electronic or magnetic level or at both levels. However, further experiments are required to confirm this.

In \emph{SIO-anneal} sample, there is a cross over from Arrhenius behaviour in the intermediate temperature region to Mott's VRH mechanism at low temperatures, Fig.9(f). As observed in \emph{SIO-as prep} sample, it is possible to model the resistivity variation by using both 2D and 3D VRH limits in the temperature range from $\sim$36K to 8K. A value of $\rho_0$=0.0042(0.2943)$\Omega$-cm and $T_M=14.14 \times 10^5(3.48 \times 10^4)$K is obtained from the fitting 3D (2D)-VRH fitting. The 3D fitting parameters obtained here matches with those reported by Kini \emph{et al}\cite{kini2006} in a similar temperature range and in polycrystalline samples. The calculated localization length $\xi$ from the 3D-VRH fitting parameters is $\sim$1.35$A^\circ$ which is even smaller than Ir-O nearest in-plane bond length. But similar to the case of \emph{SIO-as prep}, the ratio of mean hopping distance to localization length $R_M/\xi$ is greater than 1 in the temperature range of fitting which validates the use of 3D VRH model. We now investigate the effect of magnetic field on to the transport properties.

\textbf{Fig.10(a-b)} shows the percentage magnetoresistance for \emph{SIO-as prep} and \emph{SIO-anneal} samples at few selected temperatures, MR = $100 \times \rho(H)-\rho(0)/\rho(0)$, where $\rho(H)$ is the resistivity in the presence of applied magnetic field H and $\rho(0)$ is the resistivity at H=0T. In Fig.10(c), an expanded view of MR at low field is shown for \emph{SIO-as prep} sample and Fig.10(d) shows the same for \emph{SIO-anneal} sample. Both the samples exhibit negative MR whose magnitude increases with the decrease in temperature. At 300K, both the samples exhibit a negligible MR and as the temperature is lowered, there is an enhancement. At 180K and 150K, the field dependence of the MR is similar where the MR decreases sharply at low fields until $\sim$0.3T and the rate of decrease slows down towards high field. On the other hand, at 10K, MR exhibits small positive values in low fields upto $\sim$0.3T and then decreases towards high field in \emph{SIO-as prep} sample, Fig.10(c). For \emph{SIO-anneal}, MR at 10K decreases slowly in low fields, Fig.9(d), and then rapidly drops at high fields. The magnitude of MR at 9T in \emph{SIO-as prep}(\emph{SIO-anneal}) sample are $\sim$2.5$\%$($\sim$6.1$\%$), $\sim$3.7$\%$($\sim$8$\%$) and $\sim$10$\%$($\sim$18$\%$), respectively at 180K, 150K and 10K. The observed values of MR in \emph{SIO-anneal} are higher than the previously reported values for polycrystalline samples, while MR of \emph{SIO-as prep} sample are similar to the literature.\citep{bhatti2014}
\begin{figure}
\hspace{-5mm}
\includegraphics[width=0.5\textwidth]{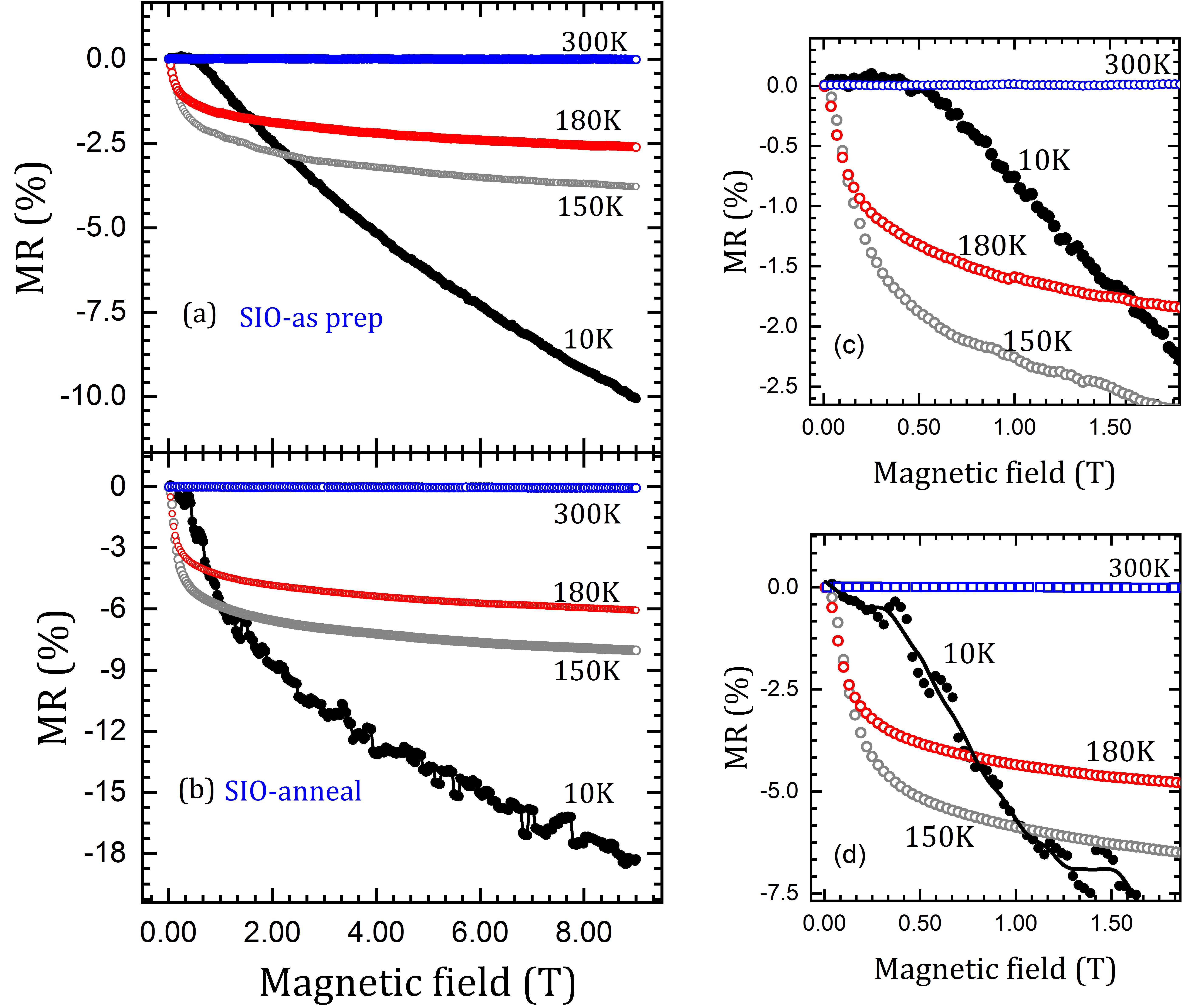}
%\vspace{-50mm}
\caption{Magnetoresistance(MR) upto 9T of (a) \emph{SIO-as prep} and (b) \emph{SIO-anneal} at different temperatures. The expanded view of MR at low fields in (c)\emph{SIO-as prep} (d)\emph{SIO-anneal} samples}
\end{figure}
The origin of negative MR has been attributed to reduction in spin dependent scattering and a slight increase in Ir-O-Ir bond angle by the application of magnetic field though strong magneto-elastic coupling.\cite{ge2011}

A larger bond angle favours electron hopping resulting in lower resistivity. Our results can also be explained by similar lines in both the samples. At $\sim$0.3T it is known that Sr$_2$IrO$_4$ undergoes a metamagnetic transition where ($\uparrow\downarrow\downarrow\uparrow$)arrangement of net magnetic moments in the \emph{ab}-plane changes to ($\uparrow\uparrow\uparrow\uparrow$) configuration between the adjacent IrO$_6$ octahedral layers.\cite{kim2009}This results in a sharp drop of MR at low fields. The difference in conduction mechanism has no effect on the nature of MR at 180K and 150K between the two samples. The higher magnitude of MR observed in \emph{SIO-anneal} sample can be explained based on its higher magnetization due to lower Ir-O-Ir bond angle which enhances ferromagnetic polarization of the domains and thus reduces spin dependent scattering to a larger extent compared to that in \emph{SIO-as prep} sample.\cite{ge2011, chen2015} Also, from structural analysis we have seen that c/a ratio and Ir-Ir interlayer distance is higher in \emph{SIO-anneal} sample which weakens the interlayer magnetic interaction. Hence,it is easier to increase Ir-O-Ir bond angle by the applied magnetic field resulting in enhanced electron hopping.\cite{ge2011}

The field dependence of MR that we see here is an average behaviour that has been reported for single crystal samples for MR along crystallographic a and c directions. In particular, MR along a-direction in single crystal samples at 10K shows an increasing trend at low fields and then decreases as the field strength increases. The positive MR at low fields and then becoming negative at high fields has been previously observed in spin glass systems such as AuFe and NiMn alloys,\cite{nigam1979} where a competing glassy magnetic phase and a ferromagnetic phases are present in the sample. In our case, canted AFM phase gives rise to negative MR with field dependence as observed in 180K and 150K. At 10K, there seems to be a positive component  which is dominant at low fields and decreases towards high fields in \emph{SIO-as prep} sample in addition to the negative component contributed by canted AFM phase. In the case of \emph{SIO-anneal} sample, the positive component in MR is lesser than that of \emph{SIO-as prep} and its MR at 10K behaves much similar to those of 180K and 150K. Thus magneto-transport at 10K suggests the presence of co-existing glassy and canted AFM phase in \emph{SIO-as prep} sample and this glassy magnetic phase is suppressed in \emph{SIO-anneal} sample. This is clear from Fig.7, that shows the extent of stabilisation of the coexisting phases in the case of \emph{SIO-as prep} is more as compared to the \emph{SIO-anneal} sample. A similar MR behaviour at low temperatures has been reported in La doped Sr$_2$IrO$_4$ single crystals which also shows spin-glass behaviour. Such glassiness is in line with the increase in the bifurcation of the FC and ZFC magnetization curves of the \emph{SIO-as prep} sample.

Now we attempt to establish a connection between the results obtained by the different experiments performed on both the samples. We have prepared polycrystalline samples of Sr$_2$IrO$_4$ using using solid state reaction method and have studied the effect of vacuum annealing on its structural and magneto-transport properties. The xrd results indicate that both the samples exhibit tetragonal \emph{I4$_1$/acd} crystal structure along with a presence of very small amounts of SrIrO$_3$ impurity. The compositional analysis by EDS technique indicates off-stoichiometry in the as prepared sample and the process of annealing brings the sample closer to stoichiometry. This off-stoichiometry may lead to the presence of Ir$^{+5}$ ions in the \emph{as prepared} sample which are smaller in size compared to Ir$^{+4}$ ions and are also non-magnetic. Annealed samples indicate higher value of magnetization as well as higher effective paramagnetic moment. Higher magnetization value is due to increased canting of moments caused by the smaller Ir-O2-Ir in-plane bond angle at all temperatures. On the other hand, higher effective paramagnetic moment might be due to conversion of some of the non-magnetic Ir$^{+5}$ ions into magnetic Ir${+4}$ ions by the removal of oxygen during vacuum annealing. The higher magnitude of Ir-O1 and Ir-O2 bond lengths in the annealed sample also supports this scenario.

The process of annealing has enhanced c/a ratio and preserved the unit cell volume at all temperatures. On the structural side, a-parameter exhibits a slope change while the c-parameter exhibits a kink at the magnetic ordering temperature showing a response to antiferromagnetic ordering in both the samples. In the \emph{as prepared} sample, the Ir-O2-Ir in-plane bond angle decreases until around 125K thereby enhancing the canting. The bond angle remains unchanged until 80K and below this temperature, it increases signalling the deviation from canted AFM structure.In the case of \emph{annealed} sample, there is an overall decrease in the bond angle until around 80K which suggests the dominance of canted AFM structure. In the temperature range 80 to 30K, the bond angle remains almost the same. Below 30K, the increment in the bond angle suggests another phase competing with the canted AFM structure. The higher magnitude of Ir-Ir out of plane bond distances in the \emph{annealed} sample indicates a weaker inter-planar coupling in comparison to the \emph{as prepared} sample. Thus the \emph{annealed} sample is expected to possess higher magnitude of magnetic anisotropy. But from the behaviour of the bifurcation in the magnetisation curves collected during FC and ZFC cycles suggests opposite behaviour for T$\leq$ 40K. Such behaviour could be due to the nature of the glassy phase stabilised at low temperatures.In the canted AFM phase, our results show the importance of the hybridisation of the Ir 5\emph{d }and O 2\emph{p} states.

The resistivity of the \emph{as prepared} sample is lower than the annealed sample at all temperatures. There are three temperature regions with a distinct transport behaviour in both the samples. From 300K to $\sim$253K, resistivity obeys Arrhenius behaviour in both the samples. From $\sim$200K to $\sim$80K,  the \emph{as prepared } sample exhibits Mott’s variable range hopping behaviour while the annealed samples show Arrhenius behavior with different activation energy compared to the previous high temperature region. Although there is an ambiguity with regard to dimensionality limit of hopping conduction, the change in conduction mechanism from VRH to Arrhenius behaviour in the intermediate temperature region is an indication of higher disorder present in the \emph{as prepared} sample compared to the annealed sample. This is also in line with the higher off-stoichiometry in the as prepared sample as observed in EDS measurements. At low temperatures, below $\sim$35K, \emph{as prepared } compound exhibits ES-VRH (T$^{-1/2}$ dependence) mechanism, while the annealed sample exhibits variable range hopping. The T$^{-1/2}$ dependence is generally observed in granular metals which contain phase separated metallic and insulating domains. This result corroborates with the field dependence of magneto resistance at 10K for the \emph{as prepared} sample which indicates a co-existing glassy and canted AFM phases. The lowest temperature magneto resistance for annealed sample indicates a suppression of the glassy magnetic phase and a dominance of canted AFM structure.This deviation in the ground state magnetic structure is also reflected in the increase in Ir-O2-Ir bond angle below $\sim$80K in the \emph{as prepared} compound. All these results suggest an electronically phase separated ground state which is probably originated by higher disorder in the \emph{as prepare}d sample.

The lowest temperature MR for annealed sample indicates a suppression of the glassy magnetic phase which is also in line with the resistivity and bond angle behaviours. The magnitude of negative MR is higher for annealed sample at all temperatures. This is due to lower Ir-O2-Ir in-plane bond angle in \emph{annealed} sample which results in higher canting of moments and hence higher value of ferromagnetic magnetization value.  At 10K, the observed magnitude of MR at 9T is around 18$\%$ which is nearly 10$\%$ higher than those reported earlier for polycrystalline sample of Sr$_2$IrO$_4$. The higher sensitivity of transport property to the applied magnetic field in the annealed samples may be useful for device applications especially in spin-orbit coupling driven AFM based spintronics.

\section{Summary}

We have explored the effect of \emph{vacuum annealing} on the magnetic, structural, transport and magneto-transport properties of polycrystalline Sr$_2$IrO$_4$ prepared by solid state reaction method. EDS analysis indicates that \emph{annealed} samples are closer to stoichiometry than the \emph{as prepared} sample. A higher effective paramagnetic moment as well as magnetization is observed in the \emph{annealed} sample. The \emph{vacuum annealing} enhances the resistivity over the entire temperature range and  both the samples exhibit distinct transport behaviour in three different temperature regions. From 300K to $\sim$253K, both the samples obey Arrhenius equation while in the intermediate temperature range from 200K $\sim$80K there is a change over from Mott's variable range hopping (VRH) to Arrhenius behaviour with the annealing. At low temperatures below $\sim$35K, the \emph{as prepared} sample obeys ES-VRH model while the \emph{annealed} samples show Mott's VRH behaviour. Our results show that vacuum annealing has increased the temperature range of stabilisation of the canted AFM phase. For both the samples we observe a signature of phase that competes with the canted AFM structure. The change in Ir-O2-Ir bond angle with annealing follow the change in the extent of magnetic anisotropy. There is an enhancement in the magnitude of magneto resistance in the annealed samples which is higher than previously reported values for the polycrystalline sample of Sr$_2$IrO$_4$. The higher magneto-resistance is explained based on larger reduction in spin dependent scattering and higher sensitivity of Ir-O2-Ir bond angle in the annealed samples. The combined analysis of magnetic, structural, transport and magneto transport results suggest a disorder induced phase separation in as prepared samples with glassy and canted AFM phases while vacuum annealing suppresses this phase separation.

\section{Acknowledgments}
Priyamedha Sharma thanks UGC-DAE CSR Indore for financial support.

%\begin{thebibliography}{99}
%
%
%
%
%\end{thebibliography}
\bibliography{sioms}

\begin{thebibliography}{45}%
\makeatletter
\providecommand \@ifxundefined [1]{%
 \@ifx{#1\undefined}
}%
\providecommand \@ifnum [1]{%
 \ifnum #1\expandafter \@firstoftwo
 \else \expandafter \@secondoftwo
 \fi
}%
\providecommand \@ifx [1]{%
 \ifx #1\expandafter \@firstoftwo
 \else \expandafter \@secondoftwo
 \fi
}%
\providecommand \natexlab [1]{#1}%
\providecommand \enquote  [1]{``#1''}%
\providecommand \bibnamefont  [1]{#1}%
\providecommand \bibfnamefont [1]{#1}%
\providecommand \citenamefont [1]{#1}%
\providecommand \href@noop [0]{\@secondoftwo}%
\providecommand \href [0]{\begingroup \@sanitize@url \@href}%
\providecommand \@href[1]{\@@startlink{#1}\@@href}%
\providecommand \@@href[1]{\endgroup#1\@@endlink}%
\providecommand \@sanitize@url [0]{\catcode `\\12\catcode `\$12\catcode
  `\&12\catcode `\#12\catcode `\^12\catcode `\_12\catcode `\%12\relax}%
\providecommand \@@startlink[1]{}%
\providecommand \@@endlink[0]{}%
\providecommand \url  [0]{\begingroup\@sanitize@url \@url }%
\providecommand \@url [1]{\endgroup\@href {#1}{\urlprefix }}%
\providecommand \urlprefix  [0]{URL }%
\providecommand \Eprint [0]{\href }%
\providecommand \doibase [0]{http://dx.doi.org/}%
\providecommand \selectlanguage [0]{\@gobble}%
\providecommand \bibinfo  [0]{\@secondoftwo}%
\providecommand \bibfield  [0]{\@secondoftwo}%
\providecommand \translation [1]{[#1]}%
\providecommand \BibitemOpen [0]{}%
\providecommand \bibitemStop [0]{}%
\providecommand \bibitemNoStop [0]{.\EOS\space}%
\providecommand \EOS [0]{\spacefactor3000\relax}%
\providecommand \BibitemShut  [1]{\csname bibitem#1\endcsname}%
\let\auto@bib@innerbib\@empty
%</preamble>
\bibitem [{\citenamefont {Witczak-Krempa}\ \emph {et~al.}(2014)\citenamefont
  {Witczak-Krempa}, \citenamefont {Chen}, \citenamefont {Kim},\ and\
  \citenamefont {Balents}}]{witczak2014}%
  \BibitemOpen
  \bibfield  {author} {\bibinfo {author} {\bibfnamefont {W.}~\bibnamefont
  {Witczak-Krempa}}, \bibinfo {author} {\bibfnamefont {G.}~\bibnamefont
  {Chen}}, \bibinfo {author} {\bibfnamefont {Y.~B.}\ \bibnamefont {Kim}}, \
  and\ \bibinfo {author} {\bibfnamefont {L.}~\bibnamefont {Balents}},\
  }\href@noop {} {\bibfield  {journal} {\bibinfo  {journal} {Annu. Rev.
  Condens. Matter Phys.}\ }\textbf {\bibinfo {volume} {5}},\ \bibinfo {pages}
  {57} (\bibinfo {year} {2014})}\BibitemShut {NoStop}%
\bibitem [{\citenamefont {Rau}\ \emph {et~al.}(2016)\citenamefont {Rau},
  \citenamefont {Lee},\ and\ \citenamefont {Kee}}]{rau2016}%
  \BibitemOpen
  \bibfield  {author} {\bibinfo {author} {\bibfnamefont {J.~G.}\ \bibnamefont
  {Rau}}, \bibinfo {author} {\bibfnamefont {E.~K.-H.}\ \bibnamefont {Lee}}, \
  and\ \bibinfo {author} {\bibfnamefont {H.-Y.}\ \bibnamefont {Kee}},\
  }\href@noop {} {\bibfield  {journal} {\bibinfo  {journal} {Annual Review of
  Condensed Matter Physics}\ }\textbf {\bibinfo {volume} {7}},\ \bibinfo
  {pages} {195} (\bibinfo {year} {2016})}\BibitemShut {NoStop}%
\bibitem [{\citenamefont {Cao}\ and\ \citenamefont
  {Schlottmann}(2018)}]{cao2018}%
  \BibitemOpen
  \bibfield  {author} {\bibinfo {author} {\bibfnamefont {G.}~\bibnamefont
  {Cao}}\ and\ \bibinfo {author} {\bibfnamefont {P.}~\bibnamefont
  {Schlottmann}},\ }\href@noop {} {\bibfield  {journal} {\bibinfo  {journal}
  {Reports on Progress in Physics}\ }\textbf {\bibinfo {volume} {81}},\
  \bibinfo {pages} {042502} (\bibinfo {year} {2018})}\BibitemShut {NoStop}%
\bibitem [{\citenamefont {Kim}\ \emph {et~al.}(2008)\citenamefont {Kim},
  \citenamefont {Jin}, \citenamefont {Moon}, \citenamefont {Kim}, \citenamefont
  {Park}, \citenamefont {Leem}, \citenamefont {Yu}, \citenamefont {Noh},
  \citenamefont {Kim}, \citenamefont {Oh} \emph {et~al.}}]{kim2008}%
  \BibitemOpen
  \bibfield  {author} {\bibinfo {author} {\bibfnamefont {B.}~\bibnamefont
  {Kim}}, \bibinfo {author} {\bibfnamefont {H.}~\bibnamefont {Jin}}, \bibinfo
  {author} {\bibfnamefont {S.}~\bibnamefont {Moon}}, \bibinfo {author}
  {\bibfnamefont {J.-Y.}\ \bibnamefont {Kim}}, \bibinfo {author} {\bibfnamefont
  {B.-G.}\ \bibnamefont {Park}}, \bibinfo {author} {\bibfnamefont
  {C.}~\bibnamefont {Leem}}, \bibinfo {author} {\bibfnamefont {J.}~\bibnamefont
  {Yu}}, \bibinfo {author} {\bibfnamefont {T.}~\bibnamefont {Noh}}, \bibinfo
  {author} {\bibfnamefont {C.}~\bibnamefont {Kim}}, \bibinfo {author}
  {\bibfnamefont {S.-J.}\ \bibnamefont {Oh}},  \emph {et~al.},\ }\href@noop {}
  {\bibfield  {journal} {\bibinfo  {journal} {Physical review letters}\
  }\textbf {\bibinfo {volume} {101}},\ \bibinfo {pages} {076402} (\bibinfo
  {year} {2008})}\BibitemShut {NoStop}%
\bibitem [{\citenamefont {Ye}\ \emph {et~al.}(2013)\citenamefont {Ye},
  \citenamefont {Chi}, \citenamefont {Chakoumakos}, \citenamefont
  {Fernandez-Baca}, \citenamefont {Qi},\ and\ \citenamefont {Cao}}]{ye2013}%
  \BibitemOpen
  \bibfield  {author} {\bibinfo {author} {\bibfnamefont {F.}~\bibnamefont
  {Ye}}, \bibinfo {author} {\bibfnamefont {S.}~\bibnamefont {Chi}}, \bibinfo
  {author} {\bibfnamefont {B.~C.}\ \bibnamefont {Chakoumakos}}, \bibinfo
  {author} {\bibfnamefont {J.~A.}\ \bibnamefont {Fernandez-Baca}}, \bibinfo
  {author} {\bibfnamefont {T.}~\bibnamefont {Qi}}, \ and\ \bibinfo {author}
  {\bibfnamefont {G.}~\bibnamefont {Cao}},\ }\href@noop {} {\bibfield
  {journal} {\bibinfo  {journal} {Physical Review B}\ }\textbf {\bibinfo
  {volume} {87}},\ \bibinfo {pages} {140406} (\bibinfo {year}
  {2013})}\BibitemShut {NoStop}%
\bibitem [{\citenamefont {Huang}\ \emph {et~al.}(1994)\citenamefont {Huang},
  \citenamefont {Soubeyroux}, \citenamefont {Chmaissem}, \citenamefont {Sora},
  \citenamefont {Santoro}, \citenamefont {Cava}, \citenamefont {Krajewski},\
  and\ \citenamefont {Peck~Jr}}]{huang1994}%
  \BibitemOpen
  \bibfield  {author} {\bibinfo {author} {\bibfnamefont {Q.}~\bibnamefont
  {Huang}}, \bibinfo {author} {\bibfnamefont {J.}~\bibnamefont {Soubeyroux}},
  \bibinfo {author} {\bibfnamefont {O.}~\bibnamefont {Chmaissem}}, \bibinfo
  {author} {\bibfnamefont {I.~N.}\ \bibnamefont {Sora}}, \bibinfo {author}
  {\bibfnamefont {A.}~\bibnamefont {Santoro}}, \bibinfo {author} {\bibfnamefont
  {R.}~\bibnamefont {Cava}}, \bibinfo {author} {\bibfnamefont {J.}~\bibnamefont
  {Krajewski}}, \ and\ \bibinfo {author} {\bibfnamefont {W.}~\bibnamefont
  {Peck~Jr}},\ }\href@noop {} {\bibfield  {journal} {\bibinfo  {journal}
  {Journal of Solid State Chemistry}\ }\textbf {\bibinfo {volume} {112}},\
  \bibinfo {pages} {355} (\bibinfo {year} {1994})}\BibitemShut {NoStop}%
\bibitem [{\citenamefont {Cava}\ \emph {et~al.}(1994)\citenamefont {Cava},
  \citenamefont {Batlogg}, \citenamefont {Kiyono}, \citenamefont {Takagi},
  \citenamefont {Krajewski}, \citenamefont {Peck~Jr}, \citenamefont {Rupp~Jr},\
  and\ \citenamefont {Chen}}]{cava1994}%
  \BibitemOpen
  \bibfield  {author} {\bibinfo {author} {\bibfnamefont {R.}~\bibnamefont
  {Cava}}, \bibinfo {author} {\bibfnamefont {B.}~\bibnamefont {Batlogg}},
  \bibinfo {author} {\bibfnamefont {K.}~\bibnamefont {Kiyono}}, \bibinfo
  {author} {\bibfnamefont {H.}~\bibnamefont {Takagi}}, \bibinfo {author}
  {\bibfnamefont {J.}~\bibnamefont {Krajewski}}, \bibinfo {author}
  {\bibfnamefont {W.}~\bibnamefont {Peck~Jr}}, \bibinfo {author} {\bibfnamefont
  {L.}~\bibnamefont {Rupp~Jr}}, \ and\ \bibinfo {author} {\bibfnamefont
  {C.}~\bibnamefont {Chen}},\ }\href@noop {} {\bibfield  {journal} {\bibinfo
  {journal} {Physical Review B}\ }\textbf {\bibinfo {volume} {49}},\ \bibinfo
  {pages} {11890} (\bibinfo {year} {1994})}\BibitemShut {NoStop}%
\bibitem [{\citenamefont {Cao}\ \emph {et~al.}(1998)\citenamefont {Cao},
  \citenamefont {Bolivar}, \citenamefont {McCall}, \citenamefont {Crow},\ and\
  \citenamefont {Guertin}}]{cao1998}%
  \BibitemOpen
  \bibfield  {author} {\bibinfo {author} {\bibfnamefont {G.}~\bibnamefont
  {Cao}}, \bibinfo {author} {\bibfnamefont {J.}~\bibnamefont {Bolivar}},
  \bibinfo {author} {\bibfnamefont {S.}~\bibnamefont {McCall}}, \bibinfo
  {author} {\bibfnamefont {J.}~\bibnamefont {Crow}}, \ and\ \bibinfo {author}
  {\bibfnamefont {R.}~\bibnamefont {Guertin}},\ }\href@noop {} {\bibfield
  {journal} {\bibinfo  {journal} {Physical Review B}\ }\textbf {\bibinfo
  {volume} {57}},\ \bibinfo {pages} {R11039} (\bibinfo {year}
  {1998})}\BibitemShut {NoStop}%
\bibitem [{\citenamefont {Wang}\ and\ \citenamefont
  {Senthil}(2011)}]{wang2011}%
  \BibitemOpen
  \bibfield  {author} {\bibinfo {author} {\bibfnamefont {F.}~\bibnamefont
  {Wang}}\ and\ \bibinfo {author} {\bibfnamefont {T.}~\bibnamefont {Senthil}},\
  }\href@noop {} {\bibfield  {journal} {\bibinfo  {journal} {Physical Review
  Letters}\ }\textbf {\bibinfo {volume} {106}},\ \bibinfo {pages} {136402}
  (\bibinfo {year} {2011})}\BibitemShut {NoStop}%
\bibitem [{\citenamefont {Yang}\ \emph {et~al.}(2014)\citenamefont {Yang},
  \citenamefont {Wang}, \citenamefont {Liu}, \citenamefont {Chen},
  \citenamefont {Dai},\ and\ \citenamefont {Wang}}]{yang2014}%
  \BibitemOpen
  \bibfield  {author} {\bibinfo {author} {\bibfnamefont {Y.}~\bibnamefont
  {Yang}}, \bibinfo {author} {\bibfnamefont {W.-S.}\ \bibnamefont {Wang}},
  \bibinfo {author} {\bibfnamefont {J.-G.}\ \bibnamefont {Liu}}, \bibinfo
  {author} {\bibfnamefont {H.}~\bibnamefont {Chen}}, \bibinfo {author}
  {\bibfnamefont {J.-H.}\ \bibnamefont {Dai}}, \ and\ \bibinfo {author}
  {\bibfnamefont {Q.-H.}\ \bibnamefont {Wang}},\ }\href@noop {} {\bibfield
  {journal} {\bibinfo  {journal} {Physical Review B}\ }\textbf {\bibinfo
  {volume} {89}},\ \bibinfo {pages} {094518} (\bibinfo {year}
  {2014})}\BibitemShut {NoStop}%
\bibitem [{\citenamefont {Gao}\ \emph {et~al.}(2015)\citenamefont {Gao},
  \citenamefont {Zhou}, \citenamefont {Huang},\ and\ \citenamefont
  {Wang}}]{gao2015}%
  \BibitemOpen
  \bibfield  {author} {\bibinfo {author} {\bibfnamefont {Y.}~\bibnamefont
  {Gao}}, \bibinfo {author} {\bibfnamefont {T.}~\bibnamefont {Zhou}}, \bibinfo
  {author} {\bibfnamefont {H.}~\bibnamefont {Huang}}, \ and\ \bibinfo {author}
  {\bibfnamefont {Q.-H.}\ \bibnamefont {Wang}},\ }\href@noop {} {\bibfield
  {journal} {\bibinfo  {journal} {Scientific reports}\ }\textbf {\bibinfo
  {volume} {5}},\ \bibinfo {pages} {1} (\bibinfo {year} {2015})}\BibitemShut
  {NoStop}%
\bibitem [{\citenamefont {Chen}\ \emph {et~al.}(2015)\citenamefont {Chen},
  \citenamefont {Hogan}, \citenamefont {Walkup}, \citenamefont {Zhou},
  \citenamefont {Pokharel}, \citenamefont {Yao}, \citenamefont {Tian},
  \citenamefont {Ward}, \citenamefont {Zhao}, \citenamefont {Parshall} \emph
  {et~al.}}]{chen2015}%
  \BibitemOpen
  \bibfield  {author} {\bibinfo {author} {\bibfnamefont {X.}~\bibnamefont
  {Chen}}, \bibinfo {author} {\bibfnamefont {T.}~\bibnamefont {Hogan}},
  \bibinfo {author} {\bibfnamefont {D.}~\bibnamefont {Walkup}}, \bibinfo
  {author} {\bibfnamefont {W.}~\bibnamefont {Zhou}}, \bibinfo {author}
  {\bibfnamefont {M.}~\bibnamefont {Pokharel}}, \bibinfo {author}
  {\bibfnamefont {M.}~\bibnamefont {Yao}}, \bibinfo {author} {\bibfnamefont
  {W.}~\bibnamefont {Tian}}, \bibinfo {author} {\bibfnamefont {T.~Z.}\
  \bibnamefont {Ward}}, \bibinfo {author} {\bibfnamefont {Y.}~\bibnamefont
  {Zhao}}, \bibinfo {author} {\bibfnamefont {D.}~\bibnamefont {Parshall}},
  \emph {et~al.},\ }\href@noop {} {\bibfield  {journal} {\bibinfo  {journal}
  {Physical Review B}\ }\textbf {\bibinfo {volume} {92}},\ \bibinfo {pages}
  {075125} (\bibinfo {year} {2015})}\BibitemShut {NoStop}%
\bibitem [{\citenamefont {De~La~Torre}\ \emph {et~al.}(2015)\citenamefont
  {De~La~Torre}, \citenamefont {Walker}, \citenamefont {Bruno}, \citenamefont
  {Ricc{\'o}}, \citenamefont {Wang}, \citenamefont {Lezama}, \citenamefont
  {Scheerer}, \citenamefont {Giriat}, \citenamefont {Jaccard}, \citenamefont
  {Berthod} \emph {et~al.}}]{de2015}%
  \BibitemOpen
  \bibfield  {author} {\bibinfo {author} {\bibfnamefont {A.}~\bibnamefont
  {De~La~Torre}}, \bibinfo {author} {\bibfnamefont {S.~M.}\ \bibnamefont
  {Walker}}, \bibinfo {author} {\bibfnamefont {F.~Y.}\ \bibnamefont {Bruno}},
  \bibinfo {author} {\bibfnamefont {S.}~\bibnamefont {Ricc{\'o}}}, \bibinfo
  {author} {\bibfnamefont {Z.}~\bibnamefont {Wang}}, \bibinfo {author}
  {\bibfnamefont {I.~G.}\ \bibnamefont {Lezama}}, \bibinfo {author}
  {\bibfnamefont {G.}~\bibnamefont {Scheerer}}, \bibinfo {author}
  {\bibfnamefont {G.}~\bibnamefont {Giriat}}, \bibinfo {author} {\bibfnamefont
  {D.}~\bibnamefont {Jaccard}}, \bibinfo {author} {\bibfnamefont
  {C.}~\bibnamefont {Berthod}},  \emph {et~al.},\ }\href@noop {} {\bibfield
  {journal} {\bibinfo  {journal} {Physical review letters}\ }\textbf {\bibinfo
  {volume} {115}},\ \bibinfo {pages} {176402} (\bibinfo {year}
  {2015})}\BibitemShut {NoStop}%
\bibitem [{\citenamefont {Pincini}\ \emph {et~al.}(2017)\citenamefont
  {Pincini}, \citenamefont {Vale}, \citenamefont {Donnerer}, \citenamefont
  {De~La~Torre}, \citenamefont {Hunter}, \citenamefont {Perry}, \citenamefont
  {Sala}, \citenamefont {Baumberger},\ and\ \citenamefont
  {McMorrow}}]{pincini2017}%
  \BibitemOpen
  \bibfield  {author} {\bibinfo {author} {\bibfnamefont {D.}~\bibnamefont
  {Pincini}}, \bibinfo {author} {\bibfnamefont {J.~G.}\ \bibnamefont {Vale}},
  \bibinfo {author} {\bibfnamefont {C.}~\bibnamefont {Donnerer}}, \bibinfo
  {author} {\bibfnamefont {A.}~\bibnamefont {De~La~Torre}}, \bibinfo {author}
  {\bibfnamefont {E.~C.}\ \bibnamefont {Hunter}}, \bibinfo {author}
  {\bibfnamefont {R.}~\bibnamefont {Perry}}, \bibinfo {author} {\bibfnamefont
  {M.~M.}\ \bibnamefont {Sala}}, \bibinfo {author} {\bibfnamefont
  {F.}~\bibnamefont {Baumberger}}, \ and\ \bibinfo {author} {\bibfnamefont
  {D.~F.}\ \bibnamefont {McMorrow}},\ }\href@noop {} {\bibfield  {journal}
  {\bibinfo  {journal} {Physical Review B}\ }\textbf {\bibinfo {volume} {96}},\
  \bibinfo {pages} {075162} (\bibinfo {year} {2017})}\BibitemShut {NoStop}%
\bibitem [{\citenamefont {Gretarsson}\ \emph {et~al.}(2016)\citenamefont
  {Gretarsson}, \citenamefont {Sung}, \citenamefont {Porras}, \citenamefont
  {Bertinshaw}, \citenamefont {Dietl}, \citenamefont {Bruin}, \citenamefont
  {Bangura}, \citenamefont {Kim}, \citenamefont {Dinnebier}, \citenamefont
  {Kim} \emph {et~al.}}]{gretarsson2016}%
  \BibitemOpen
  \bibfield  {author} {\bibinfo {author} {\bibfnamefont {H.}~\bibnamefont
  {Gretarsson}}, \bibinfo {author} {\bibfnamefont {N.}~\bibnamefont {Sung}},
  \bibinfo {author} {\bibfnamefont {J.}~\bibnamefont {Porras}}, \bibinfo
  {author} {\bibfnamefont {J.}~\bibnamefont {Bertinshaw}}, \bibinfo {author}
  {\bibfnamefont {C.}~\bibnamefont {Dietl}}, \bibinfo {author} {\bibfnamefont
  {J.~A.}\ \bibnamefont {Bruin}}, \bibinfo {author} {\bibfnamefont
  {A.}~\bibnamefont {Bangura}}, \bibinfo {author} {\bibfnamefont {Y.~K.}\
  \bibnamefont {Kim}}, \bibinfo {author} {\bibfnamefont {R.}~\bibnamefont
  {Dinnebier}}, \bibinfo {author} {\bibfnamefont {J.}~\bibnamefont {Kim}},
  \emph {et~al.},\ }\href@noop {} {\bibfield  {journal} {\bibinfo  {journal}
  {Physical review letters}\ }\textbf {\bibinfo {volume} {117}},\ \bibinfo
  {pages} {107001} (\bibinfo {year} {2016})}\BibitemShut {NoStop}%
\bibitem [{\citenamefont {Klein}\ and\ \citenamefont
  {Terasaki}(2008)}]{klein2008}%
  \BibitemOpen
  \bibfield  {author} {\bibinfo {author} {\bibfnamefont {Y.}~\bibnamefont
  {Klein}}\ and\ \bibinfo {author} {\bibfnamefont {I.}~\bibnamefont
  {Terasaki}},\ }\href@noop {} {\bibfield  {journal} {\bibinfo  {journal}
  {Journal of Physics: Condensed Matter}\ }\textbf {\bibinfo {volume} {20}},\
  \bibinfo {pages} {295201} (\bibinfo {year} {2008})}\BibitemShut {NoStop}%
\bibitem [{\citenamefont {Calder}\ \emph {et~al.}(2015)\citenamefont {Calder},
  \citenamefont {Kim}, \citenamefont {Cao}, \citenamefont {Cantoni},
  \citenamefont {May}, \citenamefont {Cao}, \citenamefont {Aczel},
  \citenamefont {Matsuda}, \citenamefont {Choi}, \citenamefont {Haskel} \emph
  {et~al.}}]{calder2015}%
  \BibitemOpen
  \bibfield  {author} {\bibinfo {author} {\bibfnamefont {S.}~\bibnamefont
  {Calder}}, \bibinfo {author} {\bibfnamefont {J.-W.}\ \bibnamefont {Kim}},
  \bibinfo {author} {\bibfnamefont {G.-X.}\ \bibnamefont {Cao}}, \bibinfo
  {author} {\bibfnamefont {C.}~\bibnamefont {Cantoni}}, \bibinfo {author}
  {\bibfnamefont {A.~F.}\ \bibnamefont {May}}, \bibinfo {author} {\bibfnamefont
  {H.~B.}\ \bibnamefont {Cao}}, \bibinfo {author} {\bibfnamefont {A.~A.}\
  \bibnamefont {Aczel}}, \bibinfo {author} {\bibfnamefont {M.}~\bibnamefont
  {Matsuda}}, \bibinfo {author} {\bibfnamefont {Y.}~\bibnamefont {Choi}},
  \bibinfo {author} {\bibfnamefont {D.}~\bibnamefont {Haskel}},  \emph
  {et~al.},\ }\href@noop {} {\bibfield  {journal} {\bibinfo  {journal}
  {Physical Review B}\ }\textbf {\bibinfo {volume} {92}},\ \bibinfo {pages}
  {165128} (\bibinfo {year} {2015})}\BibitemShut {NoStop}%
\bibitem [{\citenamefont {Lee}\ \emph {et~al.}(2012)\citenamefont {Lee},
  \citenamefont {Krockenberger}, \citenamefont {Takahashi}, \citenamefont
  {Kawasaki},\ and\ \citenamefont {Tokura}}]{lee2012}%
  \BibitemOpen
  \bibfield  {author} {\bibinfo {author} {\bibfnamefont {J.}~\bibnamefont
  {Lee}}, \bibinfo {author} {\bibfnamefont {Y.}~\bibnamefont {Krockenberger}},
  \bibinfo {author} {\bibfnamefont {K.}~\bibnamefont {Takahashi}}, \bibinfo
  {author} {\bibfnamefont {M.}~\bibnamefont {Kawasaki}}, \ and\ \bibinfo
  {author} {\bibfnamefont {Y.}~\bibnamefont {Tokura}},\ }\href@noop {}
  {\bibfield  {journal} {\bibinfo  {journal} {Physical Review B}\ }\textbf
  {\bibinfo {volume} {85}},\ \bibinfo {pages} {035101} (\bibinfo {year}
  {2012})}\BibitemShut {NoStop}%
\bibitem [{\citenamefont {Yuan}\ \emph {et~al.}(2015)\citenamefont {Yuan},
  \citenamefont {Aswartham}, \citenamefont {Terzic}, \citenamefont {Zheng},
  \citenamefont {Zhao}, \citenamefont {Schlottmann},\ and\ \citenamefont
  {Cao}}]{yuan2015}%
  \BibitemOpen
  \bibfield  {author} {\bibinfo {author} {\bibfnamefont {S.}~\bibnamefont
  {Yuan}}, \bibinfo {author} {\bibfnamefont {S.}~\bibnamefont {Aswartham}},
  \bibinfo {author} {\bibfnamefont {J.}~\bibnamefont {Terzic}}, \bibinfo
  {author} {\bibfnamefont {H.}~\bibnamefont {Zheng}}, \bibinfo {author}
  {\bibfnamefont {H.}~\bibnamefont {Zhao}}, \bibinfo {author} {\bibfnamefont
  {P.}~\bibnamefont {Schlottmann}}, \ and\ \bibinfo {author} {\bibfnamefont
  {G.}~\bibnamefont {Cao}},\ }\href@noop {} {\bibfield  {journal} {\bibinfo
  {journal} {Physical Review B}\ }\textbf {\bibinfo {volume} {92}},\ \bibinfo
  {pages} {245103} (\bibinfo {year} {2015})}\BibitemShut {NoStop}%
\bibitem [{\citenamefont {Korneta}\ \emph {et~al.}(2010)\citenamefont
  {Korneta}, \citenamefont {Qi}, \citenamefont {Chikara}, \citenamefont
  {Parkin}, \citenamefont {De~Long}, \citenamefont {Schlottmann},\ and\
  \citenamefont {Cao}}]{korneta2010}%
  \BibitemOpen
  \bibfield  {author} {\bibinfo {author} {\bibfnamefont {O.}~\bibnamefont
  {Korneta}}, \bibinfo {author} {\bibfnamefont {T.}~\bibnamefont {Qi}},
  \bibinfo {author} {\bibfnamefont {S.}~\bibnamefont {Chikara}}, \bibinfo
  {author} {\bibfnamefont {S.}~\bibnamefont {Parkin}}, \bibinfo {author}
  {\bibfnamefont {L.}~\bibnamefont {De~Long}}, \bibinfo {author} {\bibfnamefont
  {P.}~\bibnamefont {Schlottmann}}, \ and\ \bibinfo {author} {\bibfnamefont
  {G.}~\bibnamefont {Cao}},\ }\href@noop {} {\bibfield  {journal} {\bibinfo
  {journal} {Physical Review B}\ }\textbf {\bibinfo {volume} {82}},\ \bibinfo
  {pages} {115117} (\bibinfo {year} {2010})}\BibitemShut {NoStop}%
\bibitem [{\citenamefont {Wang}\ \emph {et~al.}(2018)\citenamefont {Wang},
  \citenamefont {Wang}, \citenamefont {Hu}, \citenamefont {Duan}, \citenamefont
  {Yuan}, \citenamefont {Dong}, \citenamefont {Lu},\ and\ \citenamefont
  {Liu}}]{wang2018}%
  \BibitemOpen
  \bibfield  {author} {\bibinfo {author} {\bibfnamefont {H.}~\bibnamefont
  {Wang}}, \bibinfo {author} {\bibfnamefont {W.}~\bibnamefont {Wang}}, \bibinfo
  {author} {\bibfnamefont {N.}~\bibnamefont {Hu}}, \bibinfo {author}
  {\bibfnamefont {T.}~\bibnamefont {Duan}}, \bibinfo {author} {\bibfnamefont
  {S.}~\bibnamefont {Yuan}}, \bibinfo {author} {\bibfnamefont {S.}~\bibnamefont
  {Dong}}, \bibinfo {author} {\bibfnamefont {C.}~\bibnamefont {Lu}}, \ and\
  \bibinfo {author} {\bibfnamefont {J.-M.}\ \bibnamefont {Liu}},\ }\href@noop
  {} {\bibfield  {journal} {\bibinfo  {journal} {Physical Review Applied}\
  }\textbf {\bibinfo {volume} {10}},\ \bibinfo {pages} {014025} (\bibinfo
  {year} {2018})}\BibitemShut {NoStop}%
\bibitem [{\citenamefont {Crawford}\ \emph {et~al.}(1994)\citenamefont
  {Crawford}, \citenamefont {Subramanian}, \citenamefont {Harlow},
  \citenamefont {Fernandez-Baca}, \citenamefont {Wang},\ and\ \citenamefont
  {Johnston}}]{crawford1994}%
  \BibitemOpen
  \bibfield  {author} {\bibinfo {author} {\bibfnamefont {M.}~\bibnamefont
  {Crawford}}, \bibinfo {author} {\bibfnamefont {M.}~\bibnamefont
  {Subramanian}}, \bibinfo {author} {\bibfnamefont {R.}~\bibnamefont {Harlow}},
  \bibinfo {author} {\bibfnamefont {J.}~\bibnamefont {Fernandez-Baca}},
  \bibinfo {author} {\bibfnamefont {Z.}~\bibnamefont {Wang}}, \ and\ \bibinfo
  {author} {\bibfnamefont {D.}~\bibnamefont {Johnston}},\ }\href@noop {}
  {\bibfield  {journal} {\bibinfo  {journal} {Physical Review B}\ }\textbf
  {\bibinfo {volume} {49}},\ \bibinfo {pages} {9198} (\bibinfo {year}
  {1994})}\BibitemShut {NoStop}%
\bibitem [{\citenamefont {Ranjbar}\ and\ \citenamefont
  {Kennedy}(2015)}]{ranjbar2015}%
  \BibitemOpen
  \bibfield  {author} {\bibinfo {author} {\bibfnamefont {B.}~\bibnamefont
  {Ranjbar}}\ and\ \bibinfo {author} {\bibfnamefont {B.~J.}\ \bibnamefont
  {Kennedy}},\ }\href@noop {} {\bibfield  {journal} {\bibinfo  {journal}
  {Journal of solid state chemistry}\ }\textbf {\bibinfo {volume} {232}},\
  \bibinfo {pages} {178} (\bibinfo {year} {2015})}\BibitemShut {NoStop}%
\bibitem [{\citenamefont {Dzyaloshinsky}(1958)}]{dzyaloshinsky1958}%
  \BibitemOpen
  \bibfield  {author} {\bibinfo {author} {\bibfnamefont {I.}~\bibnamefont
  {Dzyaloshinsky}},\ }\href@noop {} {\bibfield  {journal} {\bibinfo  {journal}
  {Journal of physics and chemistry of solids}\ }\textbf {\bibinfo {volume}
  {4}},\ \bibinfo {pages} {241} (\bibinfo {year} {1958})}\BibitemShut {NoStop}%
\bibitem [{\citenamefont {Jackeli}\ and\ \citenamefont
  {Khaliullin}(2009)}]{jackeli2009}%
  \BibitemOpen
  \bibfield  {author} {\bibinfo {author} {\bibfnamefont {G.}~\bibnamefont
  {Jackeli}}\ and\ \bibinfo {author} {\bibfnamefont {G.}~\bibnamefont
  {Khaliullin}},\ }\href@noop {} {\bibfield  {journal} {\bibinfo  {journal}
  {Physical review letters}\ }\textbf {\bibinfo {volume} {102}},\ \bibinfo
  {pages} {017205} (\bibinfo {year} {2009})}\BibitemShut {NoStop}%
\bibitem [{\citenamefont {Katukuri}\ \emph {et~al.}(2014)\citenamefont
  {Katukuri}, \citenamefont {Yushankhai}, \citenamefont {Siurakshina},
  \citenamefont {Van Den~Brink}, \citenamefont {Hozoi},\ and\ \citenamefont
  {Rousochatzakis}}]{katukuri2014}%
  \BibitemOpen
  \bibfield  {author} {\bibinfo {author} {\bibfnamefont {V.~M.}\ \bibnamefont
  {Katukuri}}, \bibinfo {author} {\bibfnamefont {V.}~\bibnamefont
  {Yushankhai}}, \bibinfo {author} {\bibfnamefont {L.}~\bibnamefont
  {Siurakshina}}, \bibinfo {author} {\bibfnamefont {J.}~\bibnamefont {Van
  Den~Brink}}, \bibinfo {author} {\bibfnamefont {L.}~\bibnamefont {Hozoi}}, \
  and\ \bibinfo {author} {\bibfnamefont {I.}~\bibnamefont {Rousochatzakis}},\
  }\href@noop {} {\bibfield  {journal} {\bibinfo  {journal} {Physical Review
  X}\ }\textbf {\bibinfo {volume} {4}},\ \bibinfo {pages} {021051} (\bibinfo
  {year} {2014})}\BibitemShut {NoStop}%
\bibitem [{\citenamefont {Rocha-Rodrigues}\ \emph {et~al.}(2020)\citenamefont
  {Rocha-Rodrigues}, \citenamefont {Santos}, \citenamefont {Oliveira},
  \citenamefont {Leal}, \citenamefont {Miranda}, \citenamefont {dos Santos},
  \citenamefont {Correia}, \citenamefont {Assali}, \citenamefont {Petrilli},
  \citenamefont {Araujo} \emph {et~al.}}]{rocha2020}%
  \BibitemOpen
  \bibfield  {author} {\bibinfo {author} {\bibfnamefont {P.}~\bibnamefont
  {Rocha-Rodrigues}}, \bibinfo {author} {\bibfnamefont {S.~S.}\ \bibnamefont
  {Santos}}, \bibinfo {author} {\bibfnamefont {G.~N.}\ \bibnamefont
  {Oliveira}}, \bibinfo {author} {\bibfnamefont {T.}~\bibnamefont {Leal}},
  \bibinfo {author} {\bibfnamefont {I.~P.}\ \bibnamefont {Miranda}}, \bibinfo
  {author} {\bibfnamefont {A.~M.}\ \bibnamefont {dos Santos}}, \bibinfo
  {author} {\bibfnamefont {J.~G.}\ \bibnamefont {Correia}}, \bibinfo {author}
  {\bibfnamefont {L.~V.}\ \bibnamefont {Assali}}, \bibinfo {author}
  {\bibfnamefont {H.~M.}\ \bibnamefont {Petrilli}}, \bibinfo {author}
  {\bibfnamefont {J.~P.}\ \bibnamefont {Araujo}},  \emph {et~al.},\ }\href@noop
  {} {\bibfield  {journal} {\bibinfo  {journal} {Physical Review B}\ }\textbf
  {\bibinfo {volume} {102}},\ \bibinfo {pages} {104115} (\bibinfo {year}
  {2020})}\BibitemShut {NoStop}%
\bibitem [{\citenamefont {Ge}\ \emph {et~al.}(2011)\citenamefont {Ge},
  \citenamefont {Qi}, \citenamefont {Korneta}, \citenamefont {De~Long},
  \citenamefont {Schlottmann}, \citenamefont {Crummett},\ and\ \citenamefont
  {Cao}}]{ge2011}%
  \BibitemOpen
  \bibfield  {author} {\bibinfo {author} {\bibfnamefont {M.}~\bibnamefont
  {Ge}}, \bibinfo {author} {\bibfnamefont {T.}~\bibnamefont {Qi}}, \bibinfo
  {author} {\bibfnamefont {O.}~\bibnamefont {Korneta}}, \bibinfo {author}
  {\bibfnamefont {D.}~\bibnamefont {De~Long}}, \bibinfo {author} {\bibfnamefont
  {P.}~\bibnamefont {Schlottmann}}, \bibinfo {author} {\bibfnamefont
  {W.}~\bibnamefont {Crummett}}, \ and\ \bibinfo {author} {\bibfnamefont
  {G.}~\bibnamefont {Cao}},\ }\href@noop {} {\bibfield  {journal} {\bibinfo
  {journal} {Physical Review B}\ }\textbf {\bibinfo {volume} {84}},\ \bibinfo
  {pages} {100402} (\bibinfo {year} {2011})}\BibitemShut {NoStop}%
\bibitem [{\citenamefont {Chikara}\ \emph {et~al.}(2009)\citenamefont
  {Chikara}, \citenamefont {Korneta}, \citenamefont {Crummett}, \citenamefont
  {DeLong}, \citenamefont {Schlottmann},\ and\ \citenamefont
  {Cao}}]{chikara2009}%
  \BibitemOpen
  \bibfield  {author} {\bibinfo {author} {\bibfnamefont {S.}~\bibnamefont
  {Chikara}}, \bibinfo {author} {\bibfnamefont {O.}~\bibnamefont {Korneta}},
  \bibinfo {author} {\bibfnamefont {W.}~\bibnamefont {Crummett}}, \bibinfo
  {author} {\bibfnamefont {L.}~\bibnamefont {DeLong}}, \bibinfo {author}
  {\bibfnamefont {P.}~\bibnamefont {Schlottmann}}, \ and\ \bibinfo {author}
  {\bibfnamefont {G.}~\bibnamefont {Cao}},\ }\href@noop {} {\bibfield
  {journal} {\bibinfo  {journal} {Physical Review B}\ }\textbf {\bibinfo
  {volume} {80}},\ \bibinfo {pages} {140407} (\bibinfo {year}
  {2009})}\BibitemShut {NoStop}%
\bibitem [{\citenamefont {Manna}\ \emph {et~al.}(2014)\citenamefont {Manna},
  \citenamefont {Thamizhavel},\ and\ \citenamefont {Ramakrishnan}}]{manna2014}%
  \BibitemOpen
  \bibfield  {author} {\bibinfo {author} {\bibfnamefont {P.}~\bibnamefont
  {Manna}}, \bibinfo {author} {\bibfnamefont {A.}~\bibnamefont {Thamizhavel}},
  \ and\ \bibinfo {author} {\bibfnamefont {S.}~\bibnamefont {Ramakrishnan}},\
  }in\ \href@noop {} {\emph {\bibinfo {booktitle} {Journal of Physics:
  Conference Series}}},\ Vol.\ \bibinfo {volume} {568}\ (\bibinfo
  {organization} {IOP Publishing},\ \bibinfo {year} {2014})\ p.\ \bibinfo
  {pages} {042020}\BibitemShut {NoStop}%
\bibitem [{\citenamefont {Bhatti}\ \emph {et~al.}(2014)\citenamefont {Bhatti},
  \citenamefont {Rawat}, \citenamefont {Banerjee},\ and\ \citenamefont
  {Pramanik}}]{bhatti2014}%
  \BibitemOpen
  \bibfield  {author} {\bibinfo {author} {\bibfnamefont {I.~N.}\ \bibnamefont
  {Bhatti}}, \bibinfo {author} {\bibfnamefont {R.}~\bibnamefont {Rawat}},
  \bibinfo {author} {\bibfnamefont {A.}~\bibnamefont {Banerjee}}, \ and\
  \bibinfo {author} {\bibfnamefont {A.}~\bibnamefont {Pramanik}},\ }\href@noop
  {} {\bibfield  {journal} {\bibinfo  {journal} {Journal of physics: Condensed
  matter}\ }\textbf {\bibinfo {volume} {27}},\ \bibinfo {pages} {016005}
  (\bibinfo {year} {2014})}\BibitemShut {NoStop}%
\bibitem [{\citenamefont {Kini}\ \emph {et~al.}(2006)\citenamefont {Kini},
  \citenamefont {Strydom}, \citenamefont {Jeevan}, \citenamefont {Geibel},\
  and\ \citenamefont {Ramakrishnan}}]{kini2006}%
  \BibitemOpen
  \bibfield  {author} {\bibinfo {author} {\bibfnamefont {N.}~\bibnamefont
  {Kini}}, \bibinfo {author} {\bibfnamefont {A.}~\bibnamefont {Strydom}},
  \bibinfo {author} {\bibfnamefont {H.}~\bibnamefont {Jeevan}}, \bibinfo
  {author} {\bibfnamefont {C.}~\bibnamefont {Geibel}}, \ and\ \bibinfo {author}
  {\bibfnamefont {S.}~\bibnamefont {Ramakrishnan}},\ }\href@noop {} {\bibfield
  {journal} {\bibinfo  {journal} {Journal of Physics: Condensed Matter}\
  }\textbf {\bibinfo {volume} {18}},\ \bibinfo {pages} {8205} (\bibinfo {year}
  {2006})}\BibitemShut {NoStop}%
\bibitem [{\citenamefont {Lu}\ \emph {et~al.}(2014)\citenamefont {Lu},
  \citenamefont {Quindeau}, \citenamefont {Deniz}, \citenamefont {Preziosi},
  \citenamefont {Hesse},\ and\ \citenamefont {Alexe}}]{lu2014}%
  \BibitemOpen
  \bibfield  {author} {\bibinfo {author} {\bibfnamefont {C.}~\bibnamefont
  {Lu}}, \bibinfo {author} {\bibfnamefont {A.}~\bibnamefont {Quindeau}},
  \bibinfo {author} {\bibfnamefont {H.}~\bibnamefont {Deniz}}, \bibinfo
  {author} {\bibfnamefont {D.}~\bibnamefont {Preziosi}}, \bibinfo {author}
  {\bibfnamefont {D.}~\bibnamefont {Hesse}}, \ and\ \bibinfo {author}
  {\bibfnamefont {M.}~\bibnamefont {Alexe}},\ }\href@noop {} {\bibfield
  {journal} {\bibinfo  {journal} {Applied Physics Letters}\ }\textbf {\bibinfo
  {volume} {105}},\ \bibinfo {pages} {082407} (\bibinfo {year}
  {2014})}\BibitemShut {NoStop}%
\bibitem [{\citenamefont {Sleight}\ and\ \citenamefont
  {Ramirez}(2018)}]{sleight2018}%
  \BibitemOpen
  \bibfield  {author} {\bibinfo {author} {\bibfnamefont {A.~W.}\ \bibnamefont
  {Sleight}}\ and\ \bibinfo {author} {\bibfnamefont {A.~P.}\ \bibnamefont
  {Ramirez}},\ }\href@noop {} {\bibfield  {journal} {\bibinfo  {journal} {Solid
  State Communications}\ }\textbf {\bibinfo {volume} {275}},\ \bibinfo {pages}
  {12} (\bibinfo {year} {2018})}\BibitemShut {NoStop}%
\bibitem [{\citenamefont {Longo}\ \emph {et~al.}(1971)\citenamefont {Longo},
  \citenamefont {Kafalas},\ and\ \citenamefont {Arnott}}]{longo1971}%
  \BibitemOpen
  \bibfield  {author} {\bibinfo {author} {\bibfnamefont {J.}~\bibnamefont
  {Longo}}, \bibinfo {author} {\bibfnamefont {J.}~\bibnamefont {Kafalas}}, \
  and\ \bibinfo {author} {\bibfnamefont {R.}~\bibnamefont {Arnott}},\
  }\href@noop {} {\bibfield  {journal} {\bibinfo  {journal} {Journal of Solid
  State Chemistry}\ }\textbf {\bibinfo {volume} {3}},\ \bibinfo {pages} {174}
  (\bibinfo {year} {1971})}\BibitemShut {NoStop}%
\bibitem [{\citenamefont {Kim}\ \emph {et~al.}(2009)\citenamefont {Kim},
  \citenamefont {Ohsumi}, \citenamefont {Komesu}, \citenamefont {Sakai},
  \citenamefont {Morita}, \citenamefont {Takagi},\ and\ \citenamefont
  {Arima}}]{kim2009}%
  \BibitemOpen
  \bibfield  {author} {\bibinfo {author} {\bibfnamefont {B.}~\bibnamefont
  {Kim}}, \bibinfo {author} {\bibfnamefont {H.}~\bibnamefont {Ohsumi}},
  \bibinfo {author} {\bibfnamefont {T.}~\bibnamefont {Komesu}}, \bibinfo
  {author} {\bibfnamefont {S.}~\bibnamefont {Sakai}}, \bibinfo {author}
  {\bibfnamefont {T.}~\bibnamefont {Morita}}, \bibinfo {author} {\bibfnamefont
  {H.}~\bibnamefont {Takagi}}, \ and\ \bibinfo {author} {\bibfnamefont {T.-h.}\
  \bibnamefont {Arima}},\ }\href@noop {} {\bibfield  {journal} {\bibinfo
  {journal} {Science}\ }\textbf {\bibinfo {volume} {323}},\ \bibinfo {pages}
  {1329} (\bibinfo {year} {2009})}\BibitemShut {NoStop}%
\bibitem [{\citenamefont {Haskel}\ \emph {et~al.}(2020)\citenamefont {Haskel},
  \citenamefont {Fabbris}, \citenamefont {Kim}, \citenamefont {Veiga},
  \citenamefont {Mardegan}, \citenamefont {Escanhoela~Jr}, \citenamefont
  {Chikara}, \citenamefont {Struzhkin}, \citenamefont {Senthil}, \citenamefont
  {Kim} \emph {et~al.}}]{haskel2020}%
  \BibitemOpen
  \bibfield  {author} {\bibinfo {author} {\bibfnamefont {D.}~\bibnamefont
  {Haskel}}, \bibinfo {author} {\bibfnamefont {G.}~\bibnamefont {Fabbris}},
  \bibinfo {author} {\bibfnamefont {J.}~\bibnamefont {Kim}}, \bibinfo {author}
  {\bibfnamefont {L.~S.}\ \bibnamefont {Veiga}}, \bibinfo {author}
  {\bibfnamefont {J.}~\bibnamefont {Mardegan}}, \bibinfo {author}
  {\bibfnamefont {C.}~\bibnamefont {Escanhoela~Jr}}, \bibinfo {author}
  {\bibfnamefont {S.}~\bibnamefont {Chikara}}, \bibinfo {author} {\bibfnamefont
  {V.}~\bibnamefont {Struzhkin}}, \bibinfo {author} {\bibfnamefont
  {T.}~\bibnamefont {Senthil}}, \bibinfo {author} {\bibfnamefont
  {B.}~\bibnamefont {Kim}},  \emph {et~al.},\ }\href@noop {} {\bibfield
  {journal} {\bibinfo  {journal} {Physical review letters}\ }\textbf {\bibinfo
  {volume} {124}},\ \bibinfo {pages} {067201} (\bibinfo {year}
  {2020})}\BibitemShut {NoStop}%
\bibitem [{\citenamefont {Joy}\ \emph {et~al.}(1998)\citenamefont {Joy},
  \citenamefont {Kumar},\ and\ \citenamefont {Date}}]{joy1998}%
  \BibitemOpen
  \bibfield  {author} {\bibinfo {author} {\bibfnamefont {P.}~\bibnamefont
  {Joy}}, \bibinfo {author} {\bibfnamefont {P.~A.}\ \bibnamefont {Kumar}}, \
  and\ \bibinfo {author} {\bibfnamefont {S.}~\bibnamefont {Date}},\ }\href@noop
  {} {\bibfield  {journal} {\bibinfo  {journal} {Journal of physics: condensed
  matter}\ }\textbf {\bibinfo {volume} {10}},\ \bibinfo {pages} {11049}
  (\bibinfo {year} {1998})}\BibitemShut {NoStop}%
\bibitem [{\citenamefont {Mott}(1968)}]{mott1968}%
  \BibitemOpen
  \bibfield  {author} {\bibinfo {author} {\bibfnamefont {N.}~\bibnamefont
  {Mott}},\ }\href@noop {} {\bibfield  {journal} {\bibinfo  {journal} {Journal
  of Non-Crystalline Solids}\ }\textbf {\bibinfo {volume} {1}},\ \bibinfo
  {pages} {1} (\bibinfo {year} {1968})}\BibitemShut {NoStop}%
\bibitem [{\citenamefont {Rosenbaum}(1991)}]{rosenbaum1991}%
  \BibitemOpen
  \bibfield  {author} {\bibinfo {author} {\bibfnamefont {R.}~\bibnamefont
  {Rosenbaum}},\ }\href@noop {} {\bibfield  {journal} {\bibinfo  {journal}
  {Physical Review B}\ }\textbf {\bibinfo {volume} {44}},\ \bibinfo {pages}
  {3599} (\bibinfo {year} {1991})}\BibitemShut {NoStop}%
\bibitem [{\citenamefont {Pepper}\ \emph {et~al.}(1974)\citenamefont {Pepper},
  \citenamefont {Pollitt},\ and\ \citenamefont {Adkins}}]{pepper1974}%
  \BibitemOpen
  \bibfield  {author} {\bibinfo {author} {\bibfnamefont {M.}~\bibnamefont
  {Pepper}}, \bibinfo {author} {\bibfnamefont {S.}~\bibnamefont {Pollitt}}, \
  and\ \bibinfo {author} {\bibfnamefont {C.}~\bibnamefont {Adkins}},\
  }\href@noop {} {\bibfield  {journal} {\bibinfo  {journal} {Physics Letters
  A}\ }\textbf {\bibinfo {volume} {48}},\ \bibinfo {pages} {113} (\bibinfo
  {year} {1974})}\BibitemShut {NoStop}%
\bibitem [{\citenamefont {Mott}\ and\ \citenamefont {Davis}(2012)}]{mott2012}%
  \BibitemOpen
  \bibfield  {author} {\bibinfo {author} {\bibfnamefont {N.~F.}\ \bibnamefont
  {Mott}}\ and\ \bibinfo {author} {\bibfnamefont {E.~A.}\ \bibnamefont
  {Davis}},\ }\href@noop {} {\emph {\bibinfo {title} {Electronic processes in
  non-crystalline materials}}}\ (\bibinfo  {publisher} {Oxford university
  press},\ \bibinfo {year} {2012})\BibitemShut {NoStop}%
\bibitem [{\citenamefont {{\'E}fros}\ and\ \citenamefont
  {Shklovskii}(1975)}]{efros1975}%
  \BibitemOpen
  \bibfield  {author} {\bibinfo {author} {\bibfnamefont {A.~L.}\ \bibnamefont
  {{\'E}fros}}\ and\ \bibinfo {author} {\bibfnamefont {B.~I.}\ \bibnamefont
  {Shklovskii}},\ }\href@noop {} {\bibfield  {journal} {\bibinfo  {journal}
  {Journal of Physics C: Solid State Physics}\ }\textbf {\bibinfo {volume}
  {8}},\ \bibinfo {pages} {L49} (\bibinfo {year} {1975})}\BibitemShut {NoStop}%
\bibitem [{\citenamefont {Sheng}\ \emph {et~al.}(1973)\citenamefont {Sheng},
  \citenamefont {Abeles},\ and\ \citenamefont {Arie}}]{sheng1973}%
  \BibitemOpen
  \bibfield  {author} {\bibinfo {author} {\bibfnamefont {P.}~\bibnamefont
  {Sheng}}, \bibinfo {author} {\bibfnamefont {B.}~\bibnamefont {Abeles}}, \
  and\ \bibinfo {author} {\bibfnamefont {Y.}~\bibnamefont {Arie}},\ }\href@noop
  {} {\bibfield  {journal} {\bibinfo  {journal} {Physical Review Letters}\
  }\textbf {\bibinfo {volume} {31}},\ \bibinfo {pages} {44} (\bibinfo {year}
  {1973})}\BibitemShut {NoStop}%
\bibitem [{\citenamefont {Nigam}\ and\ \citenamefont
  {Majumdar}(1979)}]{nigam1979}%
  \BibitemOpen
  \bibfield  {author} {\bibinfo {author} {\bibfnamefont {A.}~\bibnamefont
  {Nigam}}\ and\ \bibinfo {author} {\bibfnamefont {A.}~\bibnamefont
  {Majumdar}},\ }\href@noop {} {\bibfield  {journal} {\bibinfo  {journal}
  {Journal of Applied Physics}\ }\textbf {\bibinfo {volume} {50}},\ \bibinfo
  {pages} {1712} (\bibinfo {year} {1979})}\BibitemShut {NoStop}%
\end{thebibliography}%
\bibliographystyle{apsrev4-1}
\end{document}